\documentclass[useAMS,usenatbib]{mn2e}
\usepackage{psfig}
\usepackage{graphicx}

\def\cbeta{$c_{\beta}$}  
\def\kms{\relax \ifmmode {\,\rm km\,s}^{-1}\else \,km\,s$^{-1}$\fi}

\def\mincir{\ \raise-2.truept\hbox{\rlap{\hbox{$\sim$}}\raise5.truept
    \hbox{$<$}\ }}
\def\magcir{\ \raise-2.truept\hbox{\rlap{\hbox{$\sim$}}\raise5.truept
    \hbox{$>$}\ }}

\def\arcmin{$'$}

\def\arcsec{\hbox{$^{\prime\prime}$}}
\def\nii{[N {\sc ii}]}
\def\hi{H~{\sc i}}
\def\hii{H~{\sc ii}}
\def\sii{[S~{\sc ii}]}

\def\oii{[O~{\sc~ii}]}

\def\heii{He~{\sc ii}}
\def\hei{He~{\sc i}}
\def\oiii{[O~{\sc iii}]}
\def\neiii{[Ne~{\sc iii}]}

\def\ha{H$\alpha$}
\def\hb{H$\beta$}
\def\hd{H$\delta$}   
\def\hg{H$\gamma$}   

\def\te{$T_e$}

\def\ne{$N_e$}

\def\n205{NGC~205}
\def\dI{dIrr}
\def\dS{dSph}

\title[PNe: NGC 205 and the mass-metallicity relation for LG dwarfs]{Planetary nebulae: the universal mass-metallicity 
relation for Local Group dwarf galaxies and the chemistry of NGC~205\thanks{Based on observations obtained at the Gemini 
Observatory, which is operated by the Association of Universities for Research in Astronomy, 
Inc., under a cooperative agreement with the NSF on behalf of the Gemini partnership.}}

\author[D. R. Gon\c calves et al.]{Denise R. Gon\c calves$^{1,2}$\thanks{E-mail:
denise@astro.ufrj.br}, Laura Magrini$^{3}$, Ana M. Teodorescu$^{4,5}$, 
Carolina M. Carneiro$^{1}$
\\
  $^{1}$ Observat\'orio do Valongo, Universidade Federal do Rio de Janeiro, Ladeira Pedro Antonio 43, 20080-090 Rio de 
  Janeiro, Brazil\\
  $^{2}$ Instituto de Astrof\'isica de Canarias (IAC), E-38200 La Laguna, Tenerife, Spain\\
  $^{3}$ INAF - Osservatorio Astrofisico di Arcetri, Largo E. Fermi 5, I-50125 Firenze, Italy\\
  $^{4}$ Consiglio Nazionale delle Ricerche, Pisa, Italy\\
  $^{5}$ Institute for Astronomy, University of Hawaii, USA\\
}

\begin{document}

\date{Accepted ?. Received ?; in original form ?}

\pagerange{\pageref{firstpage}--\pageref{lastpage}} \pubyear{2013}

\maketitle

\label{firstpage}

\begin{abstract}
Here we study  16 planetary nebulae (PNe) in the dwarf irregular galaxy NGC~205 by using GMOS@Gemini spectra 
to derive their physical and chemical parameters. The chemical patterns and evolutionary tracks for 14 of our PNe 
suggest that there are no type I PNe among them.  
These PNe have an average oxygen abundance of 12+log(O/H)=8.08$\pm$0.28, progenitor masses 
of 2-2.5 M$_{\odot}$ and thus were born  $\sim$1.0-1.7~Gyr ago. Our results are in good agreement with previous PN  
studies in \n205. The present 12+log(O/H) is combined with our previous works and with the literature to study the PN metallicity 
trends of the Local Group (LG) dwarf galaxies, in an effort to establish the PN luminosity- and mass-metallicity relations 
(LZR and MZR) for the LG dwarf irregulars (\dI s) and dwarf spheroidals (\dS s). Previous attempts to obtain such 
relations failed to provide correct conclusions because were based on limited samples (Richer \& McCall~1995; 
Gon\c calves et al.~2007).  As far as we are able to compare stellar with nebular metallicities, our MZR   
is in very good agreement with the slope of the MZR recently obtained for LG dwarf galaxies using  spectroscopic stellar metallicities (Kirby et al.~2013).  
Actually, we found that both \dI\ and \dS\ galaxies follow the same MZR, at variance with the differences claimed 
in the past.  
Moreover our MZR is also consistent with the global MZR of star-forming galaxies, which span a wider stellar mass 
range ($\sim10^6$-~$10^{11}$M$\odot$).  
\end{abstract}


\section[]{Introduction}
Dwarf galaxies are the most numerous system in the Universe, and most of them are found in galaxy groups. 
Dwarfs in the Local Group (LG) are excellent laboratories to study galaxy evolution. Planetary nebulae (PNe) are 
among the brightest resolved stars in LG dwarf galaxies.  Their strong emission line features 
allow us to study the late evolutionary stages of stars with low and intermediate masses ($\sim$1-8 M$\odot$). 
They are also tracers of  the star formation history and chemical enrichment  in an age range from $\sim$1 to 10~Gyr ago. 

In a series of previous papers (Magrini et al. 2005, Leisy et al. 2005, 2006, Gon{\c c}alves et al. 2007, Magrini 
\& Gon{\c c}alves 2009, Gon{\c c}alves et al. 2012) we have investigated the chemical properties of the emission-line 
populations of LG dwarf galaxies, observing \hii\ regions and PNe in star-forming galaxies such as dwarf 
irregulars (\dI s) and only PNe in quiescent galaxies, such as dwarf spheroidals (\dS s). 

\subsection{NGC~205 and its PN population}
In the present paper we discuss our Gemini Multi-Object Spectrographs North (GMOS-N) spectroscopic observations of the PN population in 
the dwarf galaxy NGC~205, the brightest 
early-type dwarf satellite of Andromeda  (M31). 
NGC~205 is of particular importance among the low surface brightness galaxies in the LG
because of its interesting star formation history and several indications of a tidal encounter with its massive companion 
M31. Being located at a projected distance of 42~kpc from M31, NGC~205 is one of the closest M31 satellites.
The star formation history in NGC~205 has 
been studied extensively (e.g,. Bertola et al. 1995; Davidge 2003): an old stellar population (10 Gyr; Bica, Alloin \& Schmidt 1990) 
dominates the overall stellar content and a plume of bright blue stars, later identified as star clusters (Cappellari et al. 1999), 
was found in the central region (Baade~1944; Hodge~1973). 
Davidge (2003), studying the population of asymptotic giant branch (AGB) stars, noted that the multiple episodes of star 
formation (SF) may have occurred in NGC~205's most central regions, with a time spacing compatible with its orbital 
period (Cepa \& Beckman 1988). Therefore, tidal interactions with M31 could have triggered the latest episodes of SF. 
However, recent observations with HST point towards a constant star formation rate (SFR, Monaco et al. 2009), at least during the last $\sim$300~Myr, 
as if NGC~205 was approaching M31 for the first time (see also Howley et al. 2008). 
In addition to young stars, NGC~205 also contains gas (H {\sc i} + CO; 1.5$\times$10$^6$ M$\odot$, e.g. Young \& Lo 1997). A new 
estimate of the total gas content has been provided by De Looze et al. (2012), enlarging the estimates from previous measurements, 
but confirming the problem of  missing interstellar medium mass, i.e., an inconsistency of the measured gas mass when 
contrasted with theoretical predictions of the current expected gas content in \n205\ ($>$10$^7$ M$\odot$). 
This deficiency was explained with an efficient supernovae feedback able to expel gas/dust from the inner, star-forming 
regions of \n205. The population of PNe in \n205\ was first identified by Ford et al. (1977) and later by  Ciardullo et al. (1989). 
Corradi et al. (2005) recovered 20 of the previously known candidates and added another 55 new PNe to the candidate population, some 
of them actually belonging to the halo of M31. Richer \& McCall (2008) presented chemical abundances for thirteen among the brightest  
PNe of NGC~205, based on spectroscopic data obtained at the Canada-France-Hawaii Telescope. From their chemical composition they 
argued that the PN progenitors of the galaxy are typically  low mass stars, with M$\sim$1.5 M$\odot$ or less. Having such low mass 
progenitors, they likely do not dredge up significant quantities of oxygen, and thus these PNe can be considered good tracers of 
the interstellar medium (ISM) from which they were born.  

\subsection{The mass-metallicity relation}
The second aim of the present paper is to  establish the luminosity-metallicity relation (LZR), as well as the mass-metallicity 
relation (MZR), for the 
\dS s and \dI s of the Local Group, in a fully homogeneous way. On the shoulders of the latter lies the 
key to disentangle possible differences in star-forming and non star-forming dwarf 
galaxies, as claimed long ago by Skillman et al. (1989) and Kormendy \& Djorgovski (1989).  

A serious problem for deriving the above relations is that the metallicities are measured using different 
diagnostics in star-forming and non-star forming galaxies. The metallicity of the star-forming \dI s is 
obtained by measuring the O/H of their \hii\ regions, whereas \dS s are gas and dust free --dominated by 
the old stellar population-- so, the derived value is the [Fe/H]\footnote{Brackets indicate that the 
metallicity is given with respect to the solar metallicity.} from  red giants or other bright stars. Following Mateo~(1998), 
the oxygen abundances can be converted in iron metallicity, although the assumptions about the [O/Fe] are 
very uncertain, since the latter ratio depends on the star formation history (SFH) of the galaxies 
(Gilmore \& Wyse 1991). 
Therefore, only populations present in both \dS s and \dI s types of galaxies can 
be safely applied for such a study. In this context PNe are the ideal population to measure the 
metallicity of the dwarf spheroidals and irregulars, 
as proposed by Richer \& McCall~(1995), and actually successfully applied by these authors, as well as by Gon\c calves et 
al. (2007). Moreover, using the PN population, both groups found that the \dS s LZR presented a significant offset from the 
\dI s, with PNe in \dS s having higher O/H than those of the \dI s. Gon\c calves et al. (2007) had also compared the O/H of 
\hii\ regions of a sample of \dI s, showing no important difference between these abundances. Kormendy \& Djorgovski (1989) 
had proposed an evolutionary relation among the two morphological types of dwarfs, with the \dS\ galaxies being formed through 
the removal of the gas in dwarf irregulars, either through ram pressure stripping, supernova driven winds or star formation. 
Trying to shed light on this matter, Gon\c calves et al. (2007) added their own measurement for NGC~147 to the
data collected from literature and pointed out that the LZR of dSphs did not exclude their formation from old dwarf 
irregular galaxies, but it did exclude their formation from the present time \dI s, since the differences between
their metallicities were already present in the old PNe populations of both types. The LZR offset, then, indicates a faster 
enrichment of dSphs and the different SFH for these two types of galaxies are also discussed by Grebel (2005).   

 By combining photometric and spectroscopic stellar metallicity estimates for red giant branches, Grebel et al. (2003) 
 showed the existence of an offset between the LZR of \dS s and \dI s. These relations are such that 
 \dS s have higher mean stellar metallicities for a fixed optical luminosity, as in the case of the above discussed PN LZR. 
 They also found  
 that the same offset persists when the comparison is restricted to the galaxies old stellar populations.  Grebel et al. 
 (2003; also see Koleva et al. 2013) highlighted the ‘‘transition-type dwarfs’’ --dwarfs with mixed dIrr/dSph morphologies, 
 low stellar masses, low
angular momentum, and \hi\ contents of at most a few 10$^6$~M$_{\odot}$, which closely resemble dSph’s if their 
gas were removed--, and concluded that these transition objects are likely \dS\ progenitors. However, in  more recent 
works based on stellar metallicities (Lee et al. 2008; Kirby et al. 2013) the above offset raised some questions, since the latest 
studies, more consistently, only use spectroscopic stellar metallicities. The caveat in 
Grebel et al. (2003) approach was precisely the fact that the colors of red giants are subject to the age-metallicity degeneracy 
(Salaris \& Girardi 2005; Lianou et al. 2011). In Lee et al. (2008) the metallicities of the \dI s from RGB were re-analysed, and 
showed to be 0.5 dex higher than in Grebel et al. (2003), so vanishing the previously found offset. Kirby et al. (2013) used  
spectroscopic metallicities of individual stars in seven gas-rich dwarf irregular galaxies, and
found that dIrrs obey the same mass-metallicity relation as the \dS\ satellites of both the Milky Way and M31. Moreover, their 
MZR  is roughly continuous with the stellar mass metallicity relation for galaxies as massive as M$_∗$ = 10$^{12}$~M$_{\odot}$. 

On the controversial trends given by the latest stellar LZR/MZR (Kirby et al. 2013) as compared to the PNe LZR 
(Gon\c calves et al. 2007) resides our motivation to revisit such relation from the PN population.

\subsection{Outline of the present work}
Though in the present paper we present our new spectroscopic observations of 22 candidate PNe in \n205\ 
--for which we confirm the PN status of 19--, our work's major aims can be summarised as follows.
\begin{itemize}
\item[i)] By deriving the chemical composition of dwarf galaxies from PNe (and \hii\ regions, whenever possible), so setting 
constraints to galaxy formation and evolution, we have the unique possibility to determine directly the chemical composition 
of the ISM in different epochs than the present one, such as, for example, at the time of the formation of the PN
progenitors.
\item[ii)] Studying the stellar nucleosynthesis of low- and intermediate-mass stars in different conditions and at different 
metallicity environments. 
\item[iii)] Deriving, in a homogeneous way, the metallicity --as traced, e.g.,  by the oxygen abundance--, in both dIrrs and 
dSphs. This approach allows us to derive the luminosity (and mass)-metallicity relation for all dwarf galaxies, in a fully 
homogeneous way, to disentangle possible differences in this relation for star-forming and non star-forming galaxies, as 
claimed in the past.  
\end{itemize}

The paper is structured as follows. The acquisition and reduction of the observational data (imaging and spectroscopy) is discussed 
in Section~2. The analysis and interpretation of the spectroscopy for the 19 PNe we observed are given in Section~3, where electron 
densities and temperatures, as well as ionic and total abundances are presented, for a sub-sample of these PNe. In this section also 
appears a discussion of spectroscopic results in terms of abundances patterns and the PN progenitors age and masses.  
This is followed, in Section~4, by the determination of the luminosity-metallicity relationship and the mass-metallicity relationship 
for the LG \dI s and \dS s, as given by their PN population (in sub-sections 4.1 and 4,2 respectively). 
In Section 5 we highlight the principal results of all the previous sections of the paper.

\section[]{Observations: data acquisition and reduction}

\begin{figure} 
   \centering
   \includegraphics[width=8truecm]{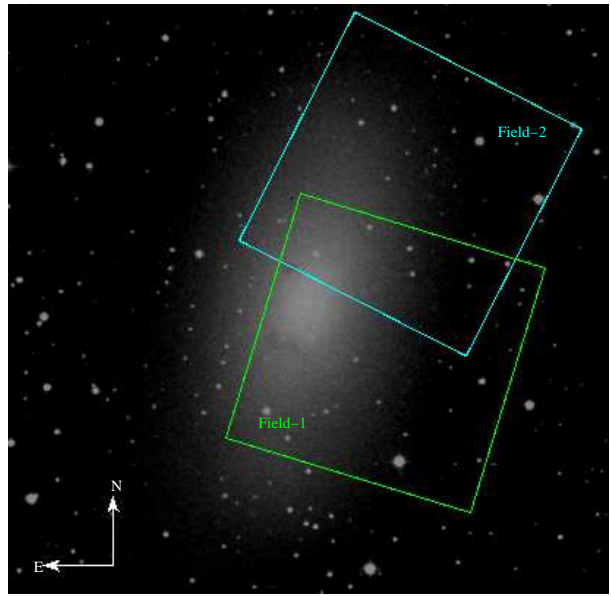} 
  \caption{A 14.2$\times$14.2~arcmin$^2$ POSSII image of NGC~205, retrieved from the NASA/IPAC Extragalactic Database.  
    The two FoV (5.5$\times$5.5~arcmin$^2$) we observed with GMOS\@Gemini are superposed to the POSSII image. 
   }
   \label{fig_fov}
\end{figure}

\begin{table}
\centering
\begin{minipage}{75mm}
{\scriptsize  
\caption{GMOS-N field identification, classification and coordinates of the \oiii\ line-emitters selected from the GMOS-N pre-imaging. F1(F2) stands for Field-1(Field-2). Our follow-up spectra give support to the object classification shown in the table.}
\begin{tabular}{@{}lllll@{}}
\hline
Field-ID & Class   & RA           & Dec  	 & C05/RM08 \\
         &         & J2000.0      & J2000.0	 &  \\        
\hline
F1-1  & star       &  00:39:57.93 & 41:40:51.06  & -       \\
F1-2  & star       &  00:40:02.16 & 41:39:35.82  & -       \\
F1-3  & star       &  00:40:06.07 & 41:37:54.33  & -       \\
F1-4  & star       &  00:40:05.55 & 41:37:37.88  & -       \\
F1-6  & star     &  00:40:05.17 & 41:39:27.32  & -	 \\
F1-7  & star     &  00:40:10.02 & 41:37:47.24  & -	 \\
F1-8  & PN &  00:40:08.77 & 41:40:43.75  & PN19/PN4 \\
F1-9  & PN &  00:40:13.47 & 41:38:42.07  & PN24/PN5 \\
F1-10 & PN &  00:40:11.07 & 41:40:48.00  & PN22 \\
F1-11 & star     &  00:40:11.37 & 41:37:48.58  & -	 \\
F1-12 & star     &  00:40:19.74 & 41:40:11.99  & -	 \\
F1-13 & PN &  00:40:20.35 & 41:38:43.80  & PN35 \\
F1-14 & PN &  00:40:17.88 & 41:38:32.85  & PN30/PN7 \\
F1-15 & PN &  00:40:21.15 & 41:38:38.76  & PN37 \\
F1-16 & PN &  00:40:20.26 & 41:38:17.55  & PN34/PN9 \\
F1-18 & -  &  00:40:20.88 & 41:41:42.10  & PN36    \\
F1-19 & star     &  00:40:23.34 & 41:40:23.02  & -	 \\
F1-20 & PN &  00:40:21.42 & 41:42:26.60  & PN38/PN1 \\
F1-21 & PN &  00:40:25.42 & 41:40:06.81  & PN42 \\
F1-22 & PN &  00:40:26.35 & 41:40:20.89  & -  \\
F2-1  & star     &  00:40:04.35 & 41:40:33.21  & -	 \\
F2-2  & star     &  00:39:55.99 & 41:43:26.19  & -	 \\
F2-3  & PN &  00:40:02.65 & 41:42:13.57  & PN16/PN19 \\
F2-4  & PN &  00:40:03.29 & 41:43:59.47  & PN17/PN24 \\
F2-5  & star     &  00:40:06.38 & 41:44:55.35  & -	 \\
F2-6  & star     &  00:40:06.10 & 41:44:41.30  & -	 \\
F2-7  & PN &  00:40:14.94 & 41:42:02.11  & PN26 \\
F2-8  & -  &  00:40:07.97 & 41:45:23.64  & PN18 \\
F2-9  & - &  00:40:17.60 & 41:41:53.30  & -    \\
F2-10 & PN &  00:40:12.00 & 41:45:31.70  & PN23 \\
F2-11 & PN &  00:40:19.17 & 41:41:47.14  & PN32 \\
F2-13 & star     &  00:40:15.72 & 41:47:06.28  & -	 \\
F2-14 & star     &  00:40:16.94 & 41:47:00.13  & -	 \\
F2-15 & PN &  00:40:17.38 & 41:45:58.63  & PN28 \\
\hline					  
\multicolumn{5}{l}{PN numbers given in column five correspond to the PN ID used } \\
\multicolumn{5}{l}{by Corradi et al. (2005, C05) and Ciardullo et al. (1989), adopted }\\ 
\multicolumn{5}{l}{by  Richer \& McCall (2008; RM08) in their spectroscopic study }\\
\multicolumn{5}{l}{of part of Ciardullo's PN candidates. Sources F1-5, F1-17 and }\\
\multicolumn{5}{l}{F2-12 where observed twice, so eliminated from this table and }\\
\multicolumn{5}{l}{further analysis. Note that F1-18, F2-8 and F2-9 are actually }\\
\multicolumn{5}{l}{symbiotic systems.}
\end{tabular}
}
\end{minipage}
\label{tab_objid}
\end{table}

We obtained pre-imaging of NGC~205 with GMOS at Gemini North telescope in 2010 observing two fields of 
5.5\arcmin$\times$5.5\arcmin\ each. In July 9 we observed Field-1 (centred at 00:40:12.50/+41:40:03) and 
September 2-3  Field-2 (00:40:10.50/+41:43:47.0). We used the on- and off-band imaging technique 
to identify PNe and other emission-line objects with  two filters: \oiii, OIII\_G0318, and \oiii-Continuum, 
OIII\_G0319, whose central $\lambda$ and width are 499~nm/5~nm, and 514~nm/10~nm, respectively. We obtained 
three exposures of 540(810)~s using the on-band(off-band) filter. We used the two combined narrow-band frames 
to build an \oiii\ continuum-subtracted image per field, from which we 
identified  a total of 37 objects to be spectroscopically  observed, including previously known PNe and other 
emission-line objects. See Table~1 
and Figure~\ref{fig_fov}. 

The spectroscopic observations were obtained in queue mode with  two gratings, R400+G5305 (`red') and B600 
(`blue') during the nights of 8 and 10 October 2010. 
The effective `blue' plus `red' spectral coverage was generally from   
3400~\AA\ to 9600~\AA, allowing an overlapping spectral range of about 300~\AA.  
Exposure times were of 3$\times$2,400~s per grating per field. 

To avoid the possibility of having important emission-lines falling in 
the gap between the 3 CCDs, we slightly varied the central wavelength of the disperser 
from one exposure to another. So we centred R400+G5305 at 750 $\pm 10$~nm and B600 at 460 $\pm 10$~nm. 

The slit width was 1\arcsec,  while the slit heights varied from 5\arcsec\ to 10\arcsec. Spatial pixels were binned. 
The final spatial scale and reciprocal dispersions of the spectra were as follows: 0\farcs161 and 0.09~nm per 
pixel, in `blue'; and 0\farcs161 and 0.134~nm per pixel, in `red'. 
The seeing varied from $\sim$0.42\arcsec\ to $\sim$0.60\arcsec. CuAr lamp exposures were obtained with both gratings, 
in the day before or after the science exposures, following the usual procedure with Gemini+GMOS for wavelength calibration.  

We performed observations of spectrophotometric standards (Massey et al. 1988; Massey \& Gronwall 1990), with the same 
instrumental setups as for science exposures. BD284211 was observed with the red grating, on October 08, 2010, whereas 
Hiltner600 was the standard for the B600, observed on October 10, 2010. 
These frames were used to flux calibrate the spectra. 

The observations were carried out at relatively low air-masses: typical airmass in \lq blue' varied from 1.08 
to 1.20; and in \lq red' from 1.27 to  1.71, even though we attempt to align the slits to the parallactic 
angle (PA), so avoiding significant losses of light, especially in the blue end of the spectra, because of 
differential atmospheric refraction. Actually, the pre-imaging was carried out with fields positioned as we can 
see from Figure~1. Then the pre-images were rotated to build the masks for spectroscopy.  For the blue exposure, 
during the observing nights the difference between the PA and the sky position angle at the telescope 
varied from 0 to 85 degree, always with air-masses above 1.2. From the Atmospheric Differential Refraction page 
in the GMOS website  {\tt http://www.gemini.edu}, we derived that at that airmass there is no light loss for 
0.42 $< \lambda <$ 0.60$\mu$m and a negligible light loss ($\leq$20\%) at $\lambda <$ 0.42$\mu$m for the highest 
difference between the PA and the sky-position angle. For the red exposures, the differences between the PA 
and the sky-position angle were $\sim$20 degree for F1 (airmass from 1.3 to 1.7), and from 25 to 65 degree for F2
(airmass from 1.4 to 1.7). Thus for F1, no light loss is expected, while for F2 we might have up to 40\% of light 
loss for $\lambda >$ 0.8$\mu$m. We double checked the three different exposures of F2 to see if significant 
amount of flux was lost in the red part of the spectra. Fortunately, the result is that the three 
exposure have similar number of counts, though they were taken over air-masses of 1.3, 1.5 and 1.7. 

Data were reduced and calibrated in the standard way by using the Gemini {\sc gmos data
  reduction script} and {\sc long-slit} tasks, both being part of
{\sc IRAF}\footnote{IRAF is distributed by the National Optical Astronomy
  Observatory, which is operated by the Association of Universities
  for Research in Astronomy (AURA) under cooperative agreement with
  the National Science Foundation.}

\section[]{Results: the spectroscopy of PNe in NGC~205}
 
\begin{figure} 
   \centering
   \includegraphics[width=8.5truecm]{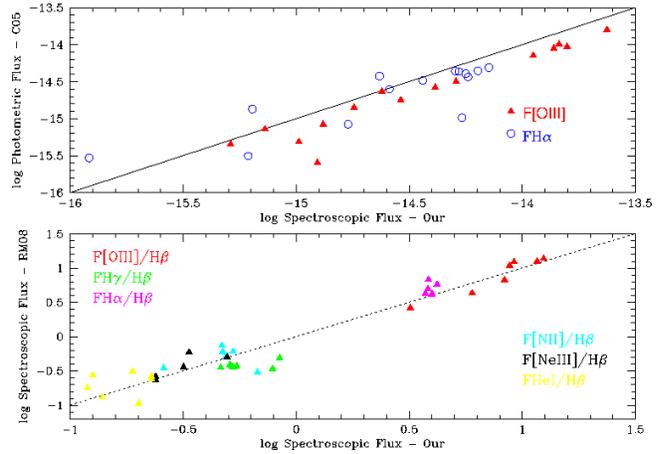} 
   \caption{Comparison of the emission-line fluxes presented in Table A1 with  the photometric fluxes 
   of C05 ({\it top panel}) and the spectroscopic fluxes of RM08 ({\it bottom panel}). Lines of different 
   wavelength and intensities are considered. 
   }
   \label{fig_flux}
\end{figure}

First, we use their spectroscopic features to classify the observed objects. 
As indicated in Table~1
,  out of the 37 slits we observed, 12 are stars; 3 are actually symbiotic systems, and all the others were 
confirmed as planetary nebulae. Some of these PNe are present in previous photometric catalogues
(Corradi et al. 2005, hereafter C05) and spectroscopic studies  (Richer \& McCall 2008, hereafter RM08). Along the 
paper, we compare our observations to the results of C05 and RM08. 

A few criteria were applied to distinguish the spectrum of a PN from  possible mimics, as done in other works 
like Pe\~na et al.~(2007) and Magrini \& Gon\c calves (2009).  
(i) PNe were selected among point-like \oiii-emitting objects. Considering \n205\ distance (0.77~Mpc; 
Howley et el.~2008), 1 arcsec corresponds to about 3.8~pc. Given the full width at half-maximum of about
0.6 arcsec for point-like objects, PNe (which have typically diameters smaller than 1~pc) are expected 
to be unresolved. 
(ii) Our PNe show faint or no continuum emission. The central stars ionising PNe are usually very hot 
(e.g. Stasi\'nska 1990; M\'endez, Kudritzki \& Herrero 1992), much more than those of the \hii\ regions. 
Due to their spectral type, with the energy maximum in the UV, PN central stars have lower M$_V$ than the 
ionising stars of \hii\ regions (typically more than 2 mag fainter). 
(iii) The presence of the \heii\ 468.6nm emission (exceeding a few percent of H${\beta}$) and/or
\oiii\ 5007/H${\beta}$ $\ge$ 4, define a spectrum as that of a PN of high-excitation. On the other hand, low-excitation 
PNe cannot be distinguished from compact \hii\ regions only on the basis of their spectroscopic line ratios.

The emission-line fluxes were measured with the package {\sc SPLOT}
of {\sc IRAF}. Errors on the fluxes were calculated taking into account
the statistical errors in the measurement of the fluxes, as well as
systematic errors (flux calibrations, background determination
and sky subtraction). The observed line fluxes were corrected for the effect of 
the interstellar extinction using the extinction law of Mathis (1990) with
R V = 3.1. We used \cbeta\ as a measurement of the extinction, which is
defined as the logarithmic difference between the observed and theoretical \hb\ fluxes. 
Since \hd\ and \hg\ are only available in few cases
and are affected by larger uncertainties, \cbeta\ was determined comparing the 
observed Balmer I(\ha)/I(\hb) ratio with its theoretical
value, 2.85 (Osterbrock \& Ferland 2006). Tables~A1, in the Appendix section, gives the results
of the emission-line flux measurements and extinction corrected
intensities for all of PNe quoted as ``PN'' in the second column of Table~1. The ID 
of the photometrically identified PN candidates listed by C05 as well as 
those whose spectroscopic analysis were presented by RM08, are also given. Note that 
all the PNe studied in these two papers, located within the FoV of Figure~1, were recovered by the 
GMOS imaging, though, because of the limited number of slits we can observe simultaneously, we 
took spectra of only 7 of the RM08 PNe. F1-18, F2-8 and F2-9 will not 
be further discussed here, but separately in a forthcoming paper, since, their spectrum show that they 
are actually symbiotic systems, instead of PNe. In the optical (Belczy\'nski et al. 2000), the 
spectra of symbiotics are indeed notable due to the absorption features and continuum of late-type M giants, the 
strong nebular emission lines of Balmer \hi, \heii\ and the forbidden lines of low- and high-ionization, like \oii, \neiii, 
etc, or those of ions with an ionization potential of at least 35 eV (e.g. \oiii). Another conspicuous discriminator 
of symbiotic systems is the presence of the Raman scattered line at $\lambda$6830\AA, a unique signature seen only in 
symbiotic stars (Schmid 1989, Belczy\'nski et al. 2000).
  
In Figure~\ref{fig_flux} we contrast the observed emission-line fluxes presented in Table~A1
with those given either from the spectroscopic data of RM08 or the photometric one of 
C05, for the objects in common. Note that we find a reasonable agreement for all the fluxes, no matter 
if photometric or spectroscopic, for a number of lines in a range of intensities and 
optical wavelengths.   

In Table~2 we show our spectroscopic results, in terms of electron densities and 
temperatures, as well as ionic and total abundances of He, O, N, Ne, Ar and S, for all the 
PNe for which these information could be extracted, i.e., a total of 14 PNe. 

\begin{table*}
\begin{minipage}{1000mm}
{\tiny  
\caption{Electron temperatures, electron densities, ionic and total abundance of the PNe.}
\begin{tabular}{@{}lrrrrrrrr@{}}
\hline
Diagnostic              &  F1-8	   	  &  F1-9	     &  F1-10	          & F1-13              &  F1-14               & F1-15		  & F1-16              & F1-20              \\
\hline
$T_{e}$[O~{\sc iii}](K)   	&  18450$\pm$400  &  13800$\pm$240   &  12850$\pm$5450    & 13700$\pm$2300     &  12350$\pm$350       & 10650$\pm$950	  & 12550$\pm$350      & 16000$\pm$2500     \\
$T_{e}$[N~{\sc ii}](K)   	&  19000$\pm$1500 &  11100$\pm$2100  &  	          &     	       &  8400$\pm$2150       & 9850$\pm$350	  & 10700$\pm$580      &	            \\
$N_{e}$~[S~{\sc ii}](cm$^{-3}$)   &  6600$\pm$50    &  4350$\pm$30     &  	          &                    &  5750$\pm$60         & 7400$\pm$200	  & 6300$\pm$60        & 3150$\pm$200 	    \\
He~{\sc i}/H 	        &  0.091       	  &  0.076	     &  0.084	          &   0.081	       &  0.084               & 0.099		  & 0.084              & 0.080              \\
He~{\sc ii}/H           &  0.015       	  &  0.006	     & -		  & -		       &  -               & -		  & -              & -              \\
He/H  	               &0.106$\pm$8.83e-04&0.082$\pm$5.93e-04& 0.084$\pm$7.59e-03 & 0.081$\pm$2.55e-03 &  0.084$\pm$0.001     & 0.099$\pm$3.67e-3 & 0.084$\pm$1.01e-3  & 0.080$\pm$05.77e-3 \\
O~{\sc i}/H 	        &  -   	  &  -       &  -	  & -		       & -		      & -	  & -          & -          \\
O~{\sc ii}/H 	        &  5.026e-06   	  &  2.511e-05       &  8.350e-05	  & 3.119e-06	       &  6.264e-05           & 1.441e-04	  & 6.588e-05          & 8.843e-06          \\
O~{\sc iii}/H 	        &  7.295e-05   	  &  1.577e-04       &  1.195e-04	  & 8.570e-05	       &  1.600e-04           & 8.807e-05	  & 1.440e-04          & 1.802e-05          \\
ICF(O)             	&  1.103       	  &  1.053	     &  1.000		  & 1.000	       &  1.000	              & 1.000		  & 1.000              & 1.000              \\
O/H  	           	&  8.606e-05   	  &  1.925e-04       &  1.195e-04	  & 8.881e-05	       &  2.226e-04           & 2.322e-04	  & 2.099e-04          & 2.686e-05          \\
12+log(O/H)      	&  7.935$\pm$0.02 &  8.284$\pm$0.10  &  8.077$\pm$0.74    & 7.948$\pm$0.204    &  8.348$\pm$0.170     & 8.366$\pm$0.105   & 8.322$\pm$0.066    & 7.429$\pm$0.205    \\
N~{\sc ii}/H  	        &  1.907e-06   	  &  6.388e-06       & -		  & 5.887e-07	       &  6.349e-06           & 1.780e-05	  & 6.225e-06          & 2.276e-06          \\
ICF(N)             	&  13.945      	  &  2.946	     & -		  & 21.359             & -		      & 1.688		  & 2.710              & 2.508              \\
N/H    	           	&  3.266e-05   	  &  4.895e-05       & -		  & 1.677e-05	       &  2.256e-05           & 2.868e-05	  & 1.983e-05          & 6.914e-06          \\
12+log(N/H)      	&  7.514$\pm$0.20  &  7.690$\pm$0.19  & -		  & 7.224$\pm$0.116    &  7.353$\pm$0.519     & 7.458$\pm$0.043   & 7.297$\pm$0.037    & 6.840$\pm$0.101    \\
Ne~{\sc iii}/H          &  8.90e-06    	  &  1.960e-05       & -		  & 8.250e-06	       &  1.740e-05           & 3.640e-06	  & 1.830e-05          & 6.760e-06          \\
ICF(Ne)            	&  1.199       	  &  1.638	     & -		  & 1.049              & -		      & 2.454		  & 1.585              & 1.663              \\
Ne/H   	           	&  1.050e-05   	  &  2.391e-05       & -		  & 8.550e-06	       &  2.421e-05           & 9.597e-06	  & 2.667e-05          & 1.008e-05          \\
12+log(Ne/H)     	&  7.021$\pm$0.006&  7.379$\pm$0.08  & -		  & 6.932$\pm$0.003    &  7.384$\pm$.136      & 6.982$\pm$0.023   & 7.426$\pm$0.032    & 7.003$\pm$0.037    \\
Ar~{\sc iii}/H          &  4.210e-07   	  &  -  	     &  4.900e-07	  & 4.390e-07	       &  7.140e-07           & 9.020e-07	  & 8.530e-07          & 5.530e-07          \\
Ar~{\sc iv}/H 	        &  -   	  &  -  	     & -		  & -		       & -		      & -	  & -          & -          \\
ICF(Ar)            	&  1.870       	  &  -  	     &  1.870		  & 1.870	       &  1.870	              & 1.870		  & 1.870              & 1.870              \\
Ar/H 	           	&  7.873e-07   	  &  -  	     &  9.163e-07	  & 8.209e-07	       &  1.335e-06           & 1.687e-06	  & 1.595e-06          & 1.034e-06          \\
12+log(Ar/H)     	&  5.896$\pm$0.37 &  -  	     & 5.962e+00$\pm$0.50 & 5.914$\pm$0.116    &  6.126$\pm$0.519     & 6.227$\pm$0.043   & 6.203$\pm$0.037    & 6.015$\pm$0.101    \\
S~{\sc ii}/H 	        &  6.130e-08   	  &  2.555e-07       & -		  & -		       &  4.015e-07           & 2.517e-07	  & 2.187e-07          & 4.678e-08          \\
S~{\sc iii}/H 	        &  1.210e-06  	  &  -       & -		  & 1.120e-06	       & -		      & -	  & -          & 2.978e-07          \\
ICF(S) 	           	&  1.710       	  &  1.120	     & -		  & 1.954              & -		      & 1.024		  & 1.101              & 1.085              \\
S/H 	           	&  2.217e-06   	  &  1.577e-06       & -		  & 2.399e-06	       &  5.441e-06           & 6.196e-06	  & 5.317e-06          & 3.884e-07          \\
12+log(S/H)      	&  6.365$\pm$0.027&  6.198$\pm$0.10  & -		  & 6.380$\pm$0.032    &  6.736$\pm$0.379     & 6.792$\pm$0.001   & 6.726$\pm$0.015    & 5.589$\pm$0.056    \\
\hline
\end{tabular}
}
\end{minipage}
\label{tabu1}
\end{table*}

\begin{table*}
\begin{minipage}{1000mm}
{\tiny  
\contcaption{}
\begin{tabular}{@{}lrrrrrr@{}}
\hline
Diagnostic            & F1-21	          &  F2-4	       & F2-7              & F2-10	       & F2-11            & F2-15	       \\ 
\hline
$T_{e}$[O~{\sc iii}](K)  & 12750$\pm$1300 &  8600$\pm$880      & 20650$\pm$1700    & 13900$\pm$1350    & 12250$\pm$650    & 11100$\pm$650      \\ 
$T_{e}$[N~{\sc ii}](K)   & - 	          &  10850$\pm$2450    & - 	           & -		       & -	          & -		       \\ 
$N_{e}$~[S~{\sc ii}](cm$^{-3}$)& 6350$\pm$200&  2650$\pm$0     & -	           & 750$\pm$30	       & -	          & -		       \\ 
He~{\sc i}/H 	      & 0.097	          &  0.145	       & -                 & 0.073	       & 0.093            & 0.083	       \\ 
He~{\sc ii}/H         & 0.004	          &  -	       & 0.062             & -	       & -            & -		       \\ 
He/H  	              & 0.101$\pm$4.32e-3 &  0.145$\pm$5.42e-3 & 0.062$\pm$8.36e-5 & 0.073$\pm$1.26e-3 & 0.093$\pm$9.89e-4& 0.083$\pm$9.00e-4 \\ 
O~{\sc i}/H 	      & -         &  -         & -	           & -         & - 	          & -		       \\ 
O~{\sc ii}/H 	      & 1.792e-05         &  9.136e-05         & -                 & 3.762e-05         & - 	          & 1.650e-06	       \\ 
O~{\sc iii}/H 	      & 8.869e-05         &  2.137e-04         & 6.875e-05         & 2.819e-05 	       & 9.420e-05        & 2.057e-04	       \\ 
ICF(O)                & 1.030	          &  1.000	       & 1.000             & 1.000	       & 1.000            & 1.000	       \\ 
O/H  	              & 1.097e-04         &  3.050e-04         & 6.875e-05         & 6.581e-05 	       & 9.420e-05        & 2.073e-04	       \\ 
12+log(O/H)           & 8.040$\pm$0.142   &  8.484$\pm$0.317   & 7.837$\pm$0.072   & 7.818$\pm$0.157   & 7.974$\pm$0.069  & 8.317$\pm$0.078    \\ 
N~{\sc ii}/H  	      & 3.093e-06         &  1.274e-05         & 8.115e-07         & 4.183e-06 	       & 4.190e-07        & 1.240e-06	       \\ 
ICF(N)                & 5.111	          &  1.751	       &  -	           & 1.611	       & - 	          & 119.458	       \\ 
N/H    	              & 1.894e-05         &  4.253e-05         &  -	           & 7.317e-06         & - 	          & 1.344e-04	       \\ 
12+log(N/H)           & 7.277$\pm$0.068   &  7.629$\pm$0.162   &  -	           & 6.864$\pm$0.071   & - 	          & 8.129e+00	       \\ 
Ne~{\sc iii}/H        & 9.820e-06         &  2.250e-05         & 8.280e-06         & - 		       & - 	          & 2.440e-05	       \\ 
ICF(Ne)               & 1.290	          &  2.331	       & 1.000             & -                 & -                & 1.008	       \\ 
Ne/H   	              & 1.215e-05         &  3.212e-05          & 8.280e-06         & - 		       & - 	          & 2.460e-05	       \\ 
12+log(Ne/H)          & 7.085$\pm$0.014   &  7.507$\pm$0.139   & 6.918$\pm$0.082   & - 		       & - 	          & 7.391$\pm$0.001    \\ 
Ar~{\sc iii}/H        & 5.470e-07         &  1.500e-06         &  -	           & 2.280e-07         & 6.350e-07        & 6.670e-07	       \\ 
Ar~{\sc iv}/H 	      & -                 &  -                 &  -	           & -         & - 	          & -		       \\ 
ICF(Ar)               & 1.870	          &  1.870	       &  -	           & 1.870	       & 1.870            & 1.870	       \\ 
Ar/H 	              & 1.023e-06         &  2.805e-06         & - 	           & 4.264e-07 	       & 1.187e-06        & 1.247e-06	       \\ 
12+log(Ar/H)          & 6.010$\pm$0.068   &  6.448$\pm$0.162   & - 	           & 5.63$\pm$0.071    & 6.075$\pm$0.031  & 6.096$\pm$0.037    \\ 
S~{\sc ii}/H 	      & 9.936e-08         &  1.698e-07         & -		   & 4.270e-08         & - 	          & -		       \\ 
S~{\sc iii}/H 	      & -                 &  8.810e-06         & - 	           & 2.140e-06 	       & 3.410e-06        & 1.000e-06	       \\ 
ICF(S) 	              & 1.278	          &  1.028	       & - 	           & 1.019	       & - 	          & 3.424	       \\ 
S/H 	              & 2.360e-06         &  1.033e-05         & - 	           & 2.243e-06 	       & - 	          & 3.482e-06	       \\ 
12+log(S/H)           & 6.373$\pm$0.019   &  7.014$\pm$0.050   & - 	           & 6.351$\pm$0.003   & - 	          & 6.542$\pm$0.006    \\ 
\hline
\end{tabular}
}
\end{minipage}       
\label{tabu2}
\end{table*}

\begin{table*}
\begin{minipage}{130mm}
{\tiny  
\caption{\cbeta, electron temperatures, electron densities and total abundances of RM08, as compared to ours, the latter 
between brackets.}
\begin{tabular}{@{}lrrrrrrr@{}}
\hline
PN ID&  \multicolumn{1}{c}{ \cbeta} & \multicolumn{1}{c}{ $T_{e}$[O~{\sc iii}]} &  \multicolumn{1}{c}{$N_{e}$~[S~{\sc ii}]} & \multicolumn{1}{c}{He/H } &  
\multicolumn{1}{c}{O/H$^a$}&  \multicolumn{1}{c}{N/H$^a$} &  \multicolumn{1}{c}{Ne/H$^a$}  \\  
		&   & \multicolumn{1}{c}{(K)}& \multicolumn{1}{c}{(cm$^{-3}$)}&    &&& \\
\hline
PN1		&  -  (0.53) & 11,500(16,000)  & 2,000$^c$(3,150) &0.131(0.080)& 8.21(7.42)&7.47(6.84)&7.48(7.00)\\
PN4		& 0.45(0.46) & 15,100(18,450)  & 2,000$^c$(6,600) &0.122(0.106)& 8.12(7.93)&8.17(7.51)&7.15(7.02)\\
PN5		& 0.42(0.36) & 11,100(13,800)  & 2,000$^c$(4,350) &0.097(0.082)& 8.53(8.28)&8.34(7.69)&7.71(7.37)\\
PN7		& 0.60(0.40) & 11,900(12,350)  & 2,000$^c$(5,750) &0.121(0.084)& 8.30(8.34)&7.87(7.35)&7.37(7.38)\\
PN9		& 0.94(0.41) & <15,200(12,550) & 2,000$^c$(6,300) &0.103(0.084)& 8.11(8.32)&7.20(7.29)&7.58(7.42)\\
PN24	        &  -  (0.12) & <16,200(8,600)  & 2,000$^c$(2,650) &0.066(0.145)& 7.55(8.48)&-	(7.62)&-   (7.50)\\
Mean$^b$        & 0.61(0.38) & 13,570(13,250)  & 2,000$^c$(5,300) &0.108(0.090)& 8.02(8.12)&7.38(7.39)& 7.33(7.22)\\
\hline
\multicolumn{8}{l}{$^a$Abundances of O, N and Ne are given in units of 12 + log(X/H).}\\ 
\multicolumn{8}{l}{$^b$The mean values are based on all the PNe described in Table~6 of RM08 and in our Table~2.} \\
\multicolumn{8}{l}{$^c$RM08 electron densities (2,000~cm$^{-3}$) are assumed, instead of observed, values.} \\
\end{tabular}
}
\end{minipage}       
\label{tabRM1}
\end{table*}

As said before, there are 7 PNe of our sample that were already studied by RM08  
(namely, in their work, as: PN1, 4, 5, 7, 9, 19 and 24). From these PNe we derived \cbeta, physical and chemical parameters 
for 6. We present in Table~3 a compilation of RM08 results, in order to make easier the comparison
of ours and their figures. 

Considering the \cbeta\ we give in Table~A1, we get a mean value of 0.38$\pm$0.09 (not taking into account the 
null cases). This mean is slightly lower than that obtained from RM08 data (0.61; the average of the 
\cbeta(0.4 R$_{\beta}$ E(B-V), R$_{\beta}$=3.041 was adopted in their paper). In Table~3 only 4 objects 
had the \cbeta\ determined using the same Balmer ratio as in our analysis. And, with the exception of PN9, 
the three other \cbeta\ values agree well. So, our \cbeta\ reasonably resembles theirs, though with a 
trend of being slightly lower. All in all, the PNe of NGC~205  have quite low extinction corrections, sometimes 
even as low as null, and never higher than 0.94 (PN9, RM08). 

From Table~\ref{tabRM1} we promptly 
note that RM08 actually did not derive the electron densities for their PNe, instead they adopted 
a fixed values of 2,000 cm$^{-3}$ to all of them.  
We, on the other hand, actually measured the ratio 
6717\AA/6737\AA\ thus, we were able to derive the \ne\sii\ for most of the PNe (see Table 2). We emphasise 
that here, for the first time, the electron density of the PNe of NGC~205 are obtained, though not 
for all PNe. As a matter of fact, with only one exception, none of these PNe has a \ne\ as low 
as the assumed value above, their mean value, with variance, being 5,300$\pm$860~cm$^{-3}$.

If electron temperatures are concerned, from our optical data it was possible to extract the \te\ based on 
the \oiii[(4959\AA+5007\AA)/4363\AA] and on the \nii[(6548\AA+6583\AA)/5755\AA] line ratios (for a number cases) 
and on one of these ratios (\oiii) for the rest of the sample. As it can be seen in Table~2, there is a good 
agreement between the two estimations of \te, if errors, instead of only the face values, are considered. 
Moreover, \te\nii/\te\oiii\ varying from 0.7 to 1.1 is expected, since they prove material at different regions 
of the nebulae and different excitation, as shown by Kaler (1986) and Krabbe \& Copetti (2005). 
On average, \te\oiii\ of our sample amounts to 13,250$\pm$3,150~K while for \te\nii\ the average value is 
12,600$\pm$4,200~K. Now, by taking a look at the comparison in Table~\ref{tabRM1}, what we get is that 
\te\oiii\ is slightly higher in our analysis than in RM08, and we do not quite know why. However, if 
instead of considering only the PNe we have in common with RM08 we look at their whole sample (their 
Table~6), the average of their  \te\oiii(variance) is 13,570($\pm$2,350~K) (this average value is computed 
avoiding the PNe for which RM08 show upper limits), which is strictly in agreement with the average value 
we got from our own data.

Finally we come to the chemical abundances of Table~2. The details of the method we use for the derivation of 
abundances are exactly the same we applied in our previous studies, like for instance in Magrini \& Gon\c calves~(2009) 
and Gon\c calves et al. (2012). In short, the {\sc ionic} task of {\sc IRAF}\footnote{The atomic data source is the that of 
{\sc analysis/nebular -- IRAF};  
http://stsdas.stsci.edu/cgi-bin/gethelp.cgi?at\_data.hlp}. 
 is 
combined with the ionisation correction 
factors from Kingsburgh \& Barlow~(1994) for the ionic and total abundances, respectively. In addition, when available, 
\te\nii\ is used for the calculation of the N$^+$, O$^+$, S$^+$ abundances,
while \te\oiii\ was used for the abundances of O$^{+2}$, S$^{+2}$, Ar$^{+2}$,
He$^+$ and He$^{+2}$. In the remaining objects, where only \te\oiii\ was
measured, we adopted it both for low- and high-ionisation species.
The abundances of \hei\ and \heii\ were computed using the equations
of Benjamin, Skillman \& Smits (1999) in two density regimes, that
is \ne\ $>$ 1,000 and $\le$ 1,000 cm$^{-3}$. The Clegg's collisional populations
were taken into account (Clegg 1987).

We discuss the figures that come up from our abundance analysis in the following sub-section --abundance 
patterns--, here we simply compare our average abundances with those of RM08. First, as a consequence of 
the significant discrepancies between our and their electron temperatures, we also find a significant 
discrepancy when comparing both abundance, if only the PNe in common are considered. This is so because 
the temperature of the gas is a strong contributor in the derivation of the ionic abundances (see, Osterbrock 
\& Ferland 2006). If, on the other hand, we take the average values of all PNe in RM08's Table~6 and our complete 
sample, both chemistries compare much better, in the last row of Table~3.

\subsection{Abundance patterns}

\begin{figure} 
   \centering
   \includegraphics[width=8.8truecm]{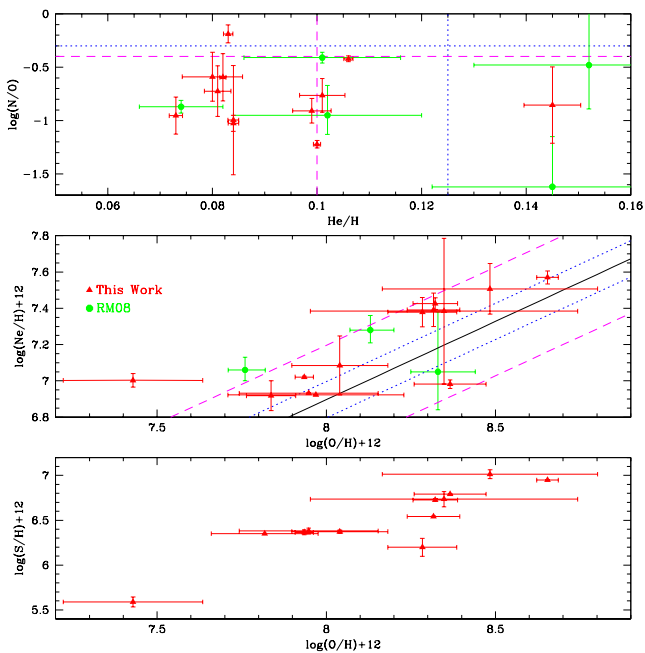} 
  \caption{Various abundance patterns. 
   {\it Top}: Log N/O vs. He/H for the brightest PNe 
   of our sample (red triangles). RM08 PNe, which were not observed in the present work 
   (see their Table~6), are 
   represented in these plots as well (green circles). He/H$\ge$0.125 and 
   log(N/O)$\ge$-0.3 give the region of the type I PNe, as defined from PNe of the Galaxy 
   Perinotto et al. (2004), marked with short dashed lines. The limits for the SMC type I PNe 
   Leisy \& Dennefeld (2006) are marked with long dashed lines. In any case type I should
   populate the top right portion of the plot. 
   {\it Middle}: Ne/H vs. O/H as compared the \lq\lq universal" relationship between these two 
   abundance ratios. The solid line is the fit for the PNe in the sample of Henry (1989). 
   The dotted lines are placed at $\pm1\sigma$ and the dashed lines at $\pm3\sigma$, where 
   $\sigma$=0.1dex.
   {\it Bottom}: S/H vs. O/H. Another $\alpha$ element, which should follow a similar 
   relationship with oxygen as that of neon, since, as much as Ne, sulphur and oxygen have 
   common origin. 
   }
   \label{fig_nhhe}
\end{figure}

Considering that we want to use the chemical compositions of the PNe to study the luminosity(mass)-metallicity 
relations of the LG dwarf galaxies of different types, the dIrrs and dSphs, we need to check if O/H we derived above is 
or not a robust diagnostic of the ISM abundances at the time the progenitor stars at the center of these PNe were born. 
Though this is a phenomenon more relevant at very low metallicities (P\'equignot et al. 2000; Leisy \& Dennefeld 
2006; Richer \& McCall 2007; Magrini \& Gon\c calves 2009) type I PNe of the low-metallicity dwarf galaxies can produce 
and dredge-up oxygen to the surface of the central star, so to the nebular gas. The O/H of these type of PNe are of 
course not a good indicator of the ISM metallicity at the time progenitors were born. The latter happening, the PN 
abundances should be used with special care in analysis such as the forthcoming ones. 

In Figure~3 we show some abundance patterns, including the log(N/O) versus He/H plot, in order to verify whether or 
not the PNe in our sample are significantly enriched in He and N. Following the definition based on Galactic PNe by 
Peimbert \& Torres-Peimbert (1983), Kingsburgh \& Barlow (1994), and others, type I PNe are nitrogen- and helium-enriched, 
with progenitors having likely undergone the third dredge-up and hot
bottom-burning, and thus are likely to have higher progenitor masses (Peimbert \& Torres-Peimbert 1983; Marigo 2001). 
By  this criterion, type I PNe are located in this plot where He/H $\ge$ 0.125 and log (N/O) $\ge$ −0.3 (short-dashed 
lines) are defined for the Milky Way (see Perinotto, Morbidelli \& Scatarzi 2004). However, because the
metallicity of \n205\ is similar to that of the Small Magellanic Cloud (SMC), we also include in this plot the equivalent 
criterion, defined by Leisy \& Dennefeld (2006) using a large number of SMC PNe (long-dashed lines). The top-right portion 
of the plot, regardless of the criterion adopted, should be populated by the type I PNe of \n205. Clearly, from these 
criteria, there are no type I objects in our, or RM08, sample of PNe in \n205. 

The other two panels of Figure~3 show the behaviour of the $\alpha$ elements (sulphur 
and neon) abundances as a function of the O/H. Neither Ne nor O are usually produced during the lifetime of 
the PN progenitors in significant amounts, and, in the case of \n205's PNe there is no reason to think 
differently. In fact, the abundances of these two elements seem to vary in lockstep, where the slope of the 
log-log relation is very close to unity (cf. Henry 1989), as we see in the middle panel of Figure~3. 
This behaviour is essentially the same for \hii\ regions (Vigroux et al. 1987; Izotov et al. 2006) and it seems 
independent of the host galaxy. As shown by Henry (1989), by putting together the Ne and O abundances of 157 PNe,  planetary 
nebulae of the Galaxy, LMC, SMC and M31 all follow the \lq\lq universal'' relation of Ne vs. O, with a 
slope of +1.16 in the log-log plot (continuous line plotted in the middle panel of the figure). S/H vs. O/H 
also follows a similar trend with oxygen as that of neon, since, sulphur and oxygen also have 
common origin. Note that RM08's PNe not re-observed in the present work follow the 
same behaviour of our own sample (green symbols). So, altogether, we can conclude that oxygen is a reliable tracer of 
the ISM composition at the time of the PN progenitor birth. 

\subsection{PN progenitors ages and masses}

\begin{figure} 
   \centering
   \includegraphics[width=8.5truecm]{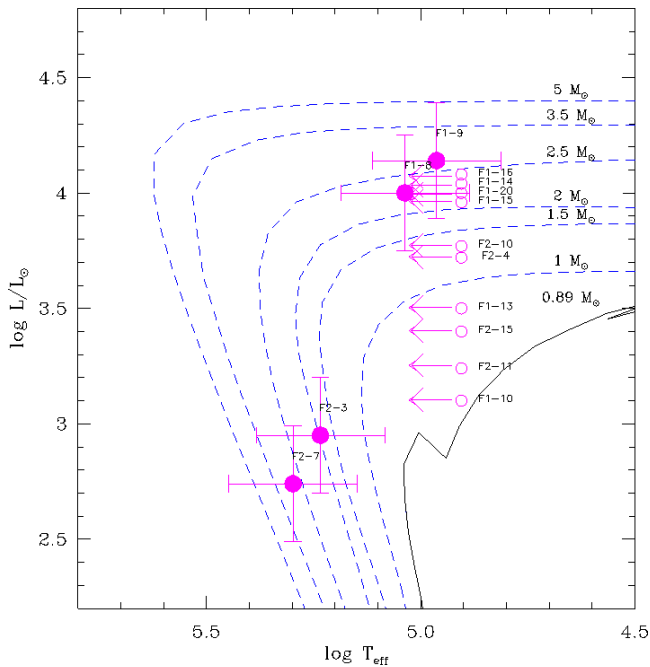} 
   \caption{Vassiliadis \& Woods (1994) tracks (z=0.004, dashed lines for H-, and continuous line for He-burning tracks), 
   with superimposed the central stars' effective temperatures 
   and luminosities, for the PNe for which we computed chemical abundances. The filled circles are PNe with \heii\ 
   detected, thus those for which the T$_{eff}$ can be directly obtained. The empty circles are PNe without \heii\ 
   lines, so their T$_{eff}$ is considered constant at 80,000 (which the value obtained in the Zanstra relationship 
   if we consider \heii~$\lambda$4686=0). 
   }
   \label{fig_nhhe}
\end{figure}

It is possible to estimate the central star luminosity and effective temperature (T$_{eff}$) 
by using our optical spectra. 
 To understand the limits and the assumptions of these estimations, there are few considerations that need to 
be done. First, T$_{eff}$ is obtained from the Zanstra temperature, using 
the formulae by Kaler \& Jacoby (1989) when applied
to extragalactic PNe  (see Kniazev et al. 2005). These relations assume that the PNe are optically thick
in the He-ionising continuum.  Thus, the derived T$_{eff}$ is an upper limit when \heii~$\lambda$4686 is not detected, 
and a lower limit when \heii~$\lambda$4686 is detected  and the nebula is optically-thin in the He-ionising continuum.  
Only in those cases when \heii~$\lambda$4686 is detected and the nebula is optically thick in the ionising 
continuum for He$^+$, so the Zanstra temperature fulfils the conditions for its derivation.  
  For the total luminosities of the PN central stars,  
the relations given in Gathier \& Pottasch (1989) and  Zijlstra \& Pottasch (1989) were applied,   
considering the photometric \oiii\ fluxes --that include both lines of the \oiii\ doublet-- of C05, and scaling them with 
the ratio between \hb\ and \oiii, to obtain the photometric \hb\ flux. 
We then corrected the \hb\ fluxes for reddening, and adopted a distance of 770~kpc to \n205 to compute the absolute luminosities. 
We note that the so-computed stellar luminosities are upper limits in all cases when the nebula is optically-thin in the H-ionising continuum. 
Masses, 
on the other hand, were derived from to theoretical evolutionary tracks of Vassiliadis \& Wood 
(1994) for Z = 0.004 (1/5 Z$_{\odot}$). Ages were estimated using the evolutionary lifetimes of
the various phases of the progenitor stars, from Vassiliadis \& Wood (1993). 
For the four PNe in which  \heii~$\lambda$4686 was  measured, we give an estimate of their luminosity and T$_{eff}$.  
For the remaining PNe, we give an upper limit of their T$_{eff}$, conservatively considering the flux of \heii~$\lambda$4686=0.  
There are two possibilities to understand their location in the HR diagram: either these stars have low temperatures, and 
they originate from very low mass progenitors --perhaps of unreasonably low mass--, or 
the detection limit for \heii\ is such that their true temperatures may be higher, near $\log$T$_{eff}$=5.2~ dex, and 
thus they are consistent with the evolutionary track of $\sim$2M${_\odot}$.  
Given the number of stars with these characteristics, the second possibility is the most probable one.
Having in mind the limits in the above calculations, we conclude that the so-derived effective temperatures and 
luminosities of the central stars 
allow us to determine an upper limit to the mass distribution for the 
progenitor stars and the age for the youngest progenitors only, which are characterised by masses of 
$\sim$2-2.5 M$_{\odot}$, and thus were born between 1.0 and 1.7~Gyr ago.

\section{The luminosity- and mass-\ metallicity relations for the LG dwarf galaxies}

\begin{figure} 
   \centering
   \includegraphics[width=8.8truecm]{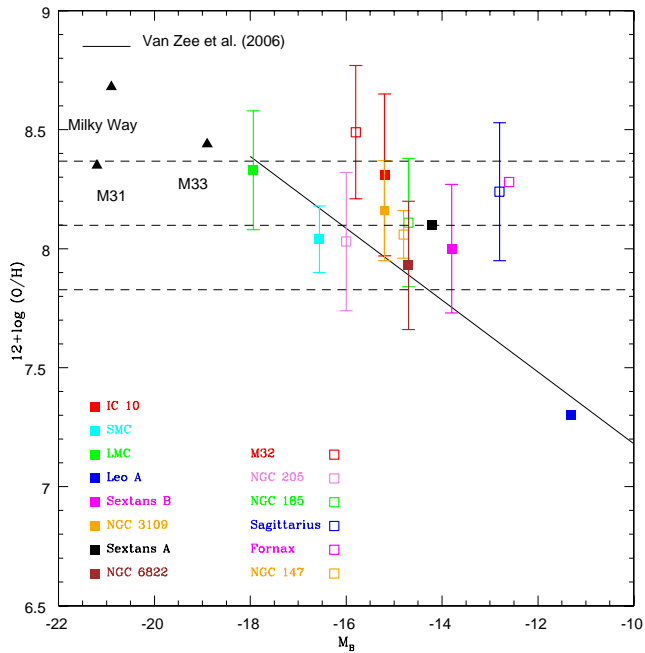} 
   \caption{The PNe luminosity-metallicity relation of the LG dwarf galaxies, showing 12+log(O/H) vs. 
   magnitude. Filled symbols represent the \dI s while the \dS s are have empty symbols. The standard 
   deviation of the abundances in a given galaxy is also plotted, except for the galaxies in which 
   only one PN had its oxygen abundance determined. The continuous line represents the LZR, from \hii\ regions 
   of nearby (D $\le$ 5~Mpc) galaxies with magnitudes fainter than -18 (van Zee et al., 2006).  The dashed 
   lines represent the mean value and standard deviation (8.098$\pm$0.27) for the PNe of all the dwarf galaxies in the plot. 
    References for the 12+log(O/H) are: IC~10, Magrini 
   \& Gon\c calves (2009); SMC, Shaw et al. (2010); LMC, Leisy \& Dennefeld (2006); Sextans~A and Sextans~B, 
   Magrini et al. (2005); Leo~A, van Zee et al. (2006); NGC~3109, Pe\~na et al. (2007); NGC~6822, 
   Hern\'andez-Martinez et al. (2009); M32, RM08; NGC~205, this work plus RM08; NGC~185, Gon\c calves et al. (2012); 
   Sagittarius, Zijlstra et al. (2006) and Otsuka et al. (2011); Fornax, Kniazev et al. (2007); and NGC~147, 
   Gon\c calves et al. (2007);  M31, Jacoby \& Ciardullo (1999), Kwitter at al. (2012), Balick et al. (2013); 
   Milky Way, Perinotto et al. (2004); M33, Magrini et al. (2009). Luminosities (M$_B$) of the dwarf galaxies are from 
   the Mateo~(1998), with the exception of those for
  the LMC and SMC that are from Lee et al.~(2003). M$_B$ of the Galaxy, M31 and M33 are 
  from van den Bergh~(2007). 
   }
   \label{fig_lm}
\end{figure}

Given the history of the LZR --the controversial existence or not of an offset in the sense that for a given luminosity 
higher metallicities would be found for the \dS s as compared with the \dI s--, we re-analyse the PN LZR and the MZR of 
the LG dwarfs, in Figure~\ref{fig_lm} and Figure~\ref{fig_mm}.

\subsection{The luminosity-metallicity relation}

To build the LZR, the oxygen abundances of the PN population of the galaxies were found in our own previous studies, and in 
a number of others from the literature (see Fig. 5). Particularly in the 
case of \n205, the mean O/H was derived from the data in the present work, the 14 PNe in Table~2, plus the 7 PNe of RM08 not 
included in this paper. Thus, the 21 PNe give a 12+log(O/H) = 8.03$\pm$0.29. 
 The luminosities (M$_B$) are from Mateo~(1998), Lee et al.~(2003) and van den Bergh~(2007). The continuous line shown in Figure~\ref{fig_lm} gives 
the weighted least-squares fit to the  metallicity versus luminosity data for the 50 \dI s within a distance of 5 Mpc. 
This fit, 
\begin{equation}
$$12 + log~({O \over H}) = 5.67 - 0.151~M_B$$,
\end{equation}
\noindent
was obtained from the galaxies' \hii\ regions, as given in the compilation of van Zee et al. (2006). The first aspect to be 
noticed in Figure~\ref{fig_lm} is that the location of the \dI s and \dS s in the LZR cannot be clearly separated from each other, 
meaning that the offset we found previously (see Fig.~6 of Gon\c calves et al. 2007) has vanished. 
The latter is due to the fact that in the present version of the plot we use new measurements of PN metallicities 
(the O/H of PN population in M32, NGC~3109, NGC~6822, NGC~185, NGC~205 and SMC come from analysis published after 2007; 
see the caption of Fig. 5). Coherently, this is in agreement with the latest results from the spectroscopically determined 
stellar LZR (Kirby et al. 2013).  
 
 We note that only one \dI\ (and none \dS) was added to the LZR since the analysis of 2007, namely IC~10 (Magrini \& 
 Gon\c calves 2009). However, the O/H of several (M32, NGC~3109, NGC~6822, NGC~185, NGC~205 and SMC) dwarf galaxies had 
 been revised, so, altogether 7 of the 14 objects in the plot had their O/H changed somewhat. 

As a matter of fact, oxygen was produced, and brought to the surface of the central stars via the third dredge-up in a 
few PNe of the \dI s Sex~A (Magrini et al. 2005) and NGC~3109 (Pe\~na et al. 2007). Judging from Figure~\ref{fig_lm}, if 
these two galaxies were excluded, the LZR of the \dI s would not change significantly. However, and fortunately, the 
phenomena of oxygen production in PNe is a rare one, and occurs mainly for 12+log(H/O)$\le$7.7 (see Magrini \& 
Gon\c calves 2009 for a complete discussion on the third dredge-up of oxygen in the LG dwarfs). 
Following Kniazev 
et al.~(2007), the self-production of oxygen occurred in the \dS\ Fornax. These authors propose to lower the 12+log(O/H) 
by 0.27$\pm$0.10 in order to reconcile the galaxy abundance patterns of S/O, Ne/O and Ar/O with the expected values. At 
variance with Sex~A and NGC~3109, in the case of Fornax, such a correction would actually make it better correlate with 
the other \dS s of the LZR. However, the O/H for Fornax is based on only one PN, so probably neither  representing the 
oxygen abundance in all of the stars in this galaxy, nor the stellar population to which its progenitor star belonged. It 
is worth mentioning as well that the 
enrichment in Sgr may have been accelerated by the lack of metal-poor gas (as compared to isolated dIrr galaxies), or 
by accretion of enriched gas expelled by our Galaxy, as pointed out in 
Zijlstra et al. (2006). A decrease of the O/H of Sgr would better locate this galaxy in the LZR defined by the remaining \dS s. 
Also, a higher luminosity for Sgr would make it better agrees with the location of the other data points in the relation, and, 
in all likelihood, its luminosity before its interaction with the Milky Way was substantially higher than that of Fornax.

\subsection{The mass-metallicity relation} 

\begin{figure} 
   \centering
   \includegraphics[width=9.0truecm]{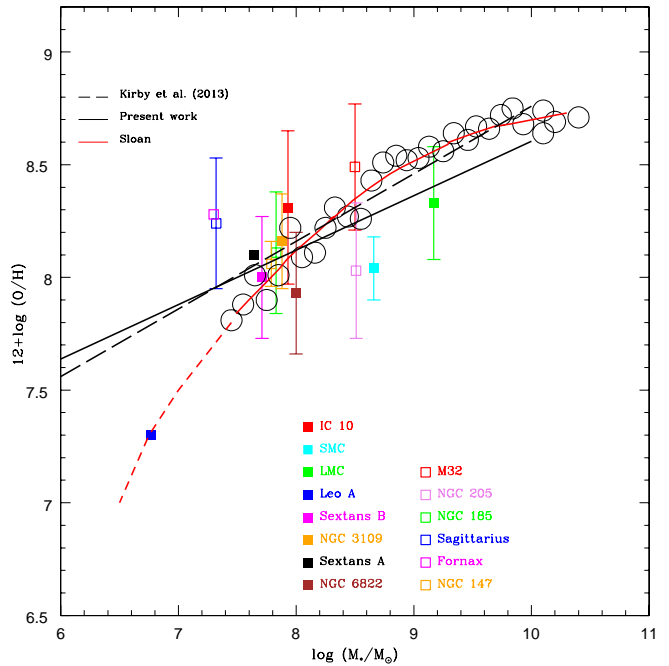} 
   \caption{The PN mass-metallicity relation of the LG dwarf galaxies, showing 12+log(O/H) vs. 
   stellar mass. Filled symbols represent the \dI s while the \dS s are have empty symbols. The standard 
   deviation of the abundances in a given galaxy is also plotted, except for the galaxies in which 
   only one PN has its oxygen abundance determined. References for the 12+log(O/H) are the same as 
   in Figure~\ref{fig_lm}. Data for stellar masses were taken from the compilation by McConnachie~(2012). 
   The dashed line represents the mass-stellar metallicity relation of Kirby et al. (2013) to which we added a 
   constant to match the oxygen metallicity, while the continuous line 
   is the mean least square fit to our data. The direct method mass-metallicity relation for M$_\star$ stacks in  
   Sloan SDSS Data Release 7 (DR7; Abazajian et al. 2009) (empty circles) from Andrews \& Martini (2013). The red 
   solid curve shows the asymptotic logarithmic fit to the direct method measurements. The dashed red curve is an 
   extrapolation of the fit.  }
   \label{fig_mm}
\end{figure}

In addition to the LZR,  we present in Figure~\ref{fig_mm} the mass-metallicity relation 
of the LG dwarf galaxies. The stellar masses of the LG galaxies were taken from the compilation by McConnachie~(2012), 
while the oxygen abundances are from the analysis of PN spectra, as in Fig.~\ref{fig_lm}. First, we note that, as in 
the case of the LZR, \dI s and \dS s are not segregated, and follow a common relation. 
 
In Figure~\ref{fig_mm} we plot the least-squares fit obtained by Kirby et al. (2013, see their eq.~4) using the stellar 
metallicities in dwarf galaxies and our least mean square fit. We added a constant to the Kirby's et al. (2013) 
interception of the Y-axis to match the nebular metallicity, expressed in our plot as 12 + log(O/H).
The Kirby's relation becomes: 
\begin{equation}
12 + log~({O \over H})=(-1.69-1.8+A)+0.3\times log~({M\star \over M_{\odot}})
\end{equation}
where -1.69 is the original intercept from the fit of Kirby et al. (2013), obtained using log~(M$\star\times10^6$M$_{\odot}$),  
$-$1.8 is the constant that takes into account that our abscissa is expressed in log(M$\star$/M$_{\odot}$)
and A=9.25 is the constant needed to match the data.  
 The choice of A=9.25 dex translates the Kirby et al. (2013) relation to the
region of Fig. 6 occupied by our data. While the slope is relevant for
comparison with our data and the other relations in Fig. 6, the zero point is
not.
The least mean square fit to our data gives the following relation: 
\begin{equation}
12 + log~({O \over H})=6.19 + 0.24\times log~({M\star \over M_{\odot}}). 
\end{equation}
 The fit presented in Eq.~3 is very sensitive to the data points for the LMC
and Leo A. However, excluding them would mean to limit the mass range from log(M$\star$/M$_{\odot}$)$\sim$7.5 to 8.

The lowest mass in Kirby's et al (2013) was of the order of 10$^3$M$_{\odot}$.  In Figure~\ref{fig_mm}, we encompass a 
smaller stellar mass range (10$^7$M$_{\odot}$  to 10$^9$M$_{\odot}$) in which we find that the MZR has a  slope similar 
to that of the relation built with stellar metallicities. The MZR with oxygen abundances seems to deviate only when 
reaching the stellar mass of the \dI\ Leo~A with M$_{\star}<$10$^7$M$_{\odot}$. 

In Figure~\ref{fig_mm}, we also show the comparison with the mass-metallicity relation for M$_\star$ stacks, obtained 
with Sloan SDSS Data Release 7 by Andrews \& Martini (2013). They  measured   \oiii, \oii, \nii, and \sii\ electron 
temperatures, and from them, they adopted the direct method gas-phase oxygen abundances from stacked galaxy spectra. 
They stacked the spectra of $\sim$200,000~SDSS star-forming galaxies in bins of  0.1 dex in stellar mass. 
Thus, the Andrews \& Martini (2013)'s MZR is better comparable with our results than MZR  obtained with  methods 
that do rely on strong line diagnostics (e.g. Tremonti  et al. 2004).  

Excluding Fornax and Sgr, discussed in the previous section, we note that the oxygen abundances of  PNe trace the 
same metallicity as {H {\sc ii}} regions in the Sloan SDSS. In particular we do not see any offset from the average 
metallicity in each bin of stellar mass. We also note that the \dI\ galaxy Leo~A deviates from the linear fit, and it 
is located below both the Van Zee et al. (2006)'s fit of Fig.~\ref{fig_lm} and the Kirby et al. (2013)'s fit of 
Fig.~\ref{fig_mm}. However, the location of Leo~A  is perfectly consistent with the extrapolation of the asymptotic 
logarithmic fit of the SLOAN mass-metallicity relation shown in Fig.~\ref{fig_mm}. 
Note that this is among the metal poorest points added to the MZR relation built with oxygen abundances. 

The results shown in Figure~\ref{fig_mm} points out a {\em possible} universality of the MZR for both \dI\ and \dS\ galaxies. 
 There is still a non-negligible  
scatter among the data points based upon the chemical
composition from PNe. 
 At variance with the previous results (see Richer \& McCall~1995, Gon\c calves et al.~2007), 
based on small statistics and non-homogeneous analysis and indicating a clearly different behaviour of  \dI\ and \dS\ galaxies, 
now  these two classes of galaxies are compatible with a unique mass-metallicity relation. 
 Being very conservative, we can conclude saying that 
at present neither the data nor the mass-metallicity relations are
known well enough to {\em categorically} affirm that dwarf irregulars and
dwarf spheroidals follow the same mass-metallicity relation, though they are
now fully compatible with this possibility. This was not the case previously. 
 It is interesting to notice that both \dI s and \dS s are consistent with global MZR of star forming galaxies, 
likely because chemical evolution is a function of stellar mass and its correlation with the total mass (baryonic and 
non-baryonic) of the galaxy.

\section{Conclusions}

In this paper our new optical observations for \n205, obtained using the GMOS at Gemini, a multi object 
imager and spectrograph, were presented. By applying the narrow-band \oiii\ - \oiii$_{\sc cont}$ technique, we identified 
a number of 37 PN candidates, which were then studied in terms of their physical and chemical properties. 16 of these candidates 
were actually confirmed to be PNe. Another three were identified as symbiotic systems (Gon\c calves et al, in prep).

The analysis of the PN optical spectra allowed us to derive the electron densities (\ne; none of the \n205\ PNe had its 
\ne\ determined previous to the present work), the electron temperatures (\te) and ionic as well as total abundances. 
These PN properties were compared with the only good enough previous work about the spectroscopy of the PNe in \n205\ 
(Richer \& McCall 2008) and, in general, both analysis are in agreement. 

The chemical patterns and evolutionary tracks derived for 14 of our PNe suggest that there are no type I planetary 
nebulae in the present sample, which also implies that the PN average oxygen abundance (12+log(O/H)=8.08$\pm$0.28) 
of \n205\ is a robust diagnostic of the ISM metallicity by the time the progenitor stars were born, about 1.0 to 1.7~Gyr 
ago. Altogether \n205's PNe are relatively young (latter figures),  as it is implied by the mass range of their progenitors, between 2 
and 2.5~M$_{\odot}$.

We have been studying for years the metallicity trends of the LG dwarf galaxies, as seen from these galaxies PN population 
(Magrini et al. 2005, Leisy et al. 2005, 2006, Gon{\c c}alves et al. 2007, Magrini \& Gon{\c c}alves 2009, Gon{\c c}alves 
et al. 2012), among others, with the goal to  establish the PNe luminosity- and mass-metallicity relations for the nearby 
\dI s as well as the \dS s. Previous attempts to obtain such relations have been shown to be based on a too limited sample 
and poorly measured abundances (Richer \& McCall 1995; Gon\c calves et al. 2007). We revisited this issue using more and 
more accurate data, and found results that contradict the previous ones, but that agree very closely with the new LZR and
MZR obtained recently by 
using  stellar metallicities (Kirby et al. 2013). 

Using spectroscopic stellar metallicities for \dI s and \dS s, Kirby et al. (2013) have recently shown that dwarf 
irregulars  obey the same mass-metallicity relation as the dwarf spheroidals, and they also have shown this relationship 
is roughly continuous with the stellar mass-stellar metallicity relation for galaxies as massive as M$_{*}$ = 10$^{12}$~M$_{\odot}$. 
It is extremely interesting to note that, the slope of the MZR given by the PNe metallicities in the present work, is 
the same as that obtained by using stellar metallicities. The stellar masses of the galaxies in our sample correspond 
to ~10$^7$ up to ~10$^9$~M$_{\odot}$. On the other hand, in Kirby et al. (2013) galaxies of smaller stellar masses 
($>$10$^{3}$~M$_{\odot}$) are also studied, and, in addition, using the \hii\ region metallicities recently obtained 
for more massive galaxies (SDSS; Andrews \& Martini (2013), the universality of the MZR is strongly reinforced by the 
present analysis. 

And, a final aspect of the our results that it is worth highlighting is that not only the MZR for both \dI\ and \dS\ 
galaxies are consistent with the same relation, at variance with the differences claimed in the past, but also the MZR of both \dI s and \dS s 
are consistent with the global MZR of star-forming galaxies. 

\section{Acknowledgments}
Authors are very grateful with Anna Gallazzi and Evan Kirby for their critical reading of the manuscript and a number of 
fruitful discussions. We also thanks the careful report of the referee, Michael Richer, which helped us to 
significantly improve the paper. 
DRG kindly acknowledges the Instituto de Astrof\'isica de Canarias (IAC) for their hospitality, where part of this
work was done. CMC is supported by CAPES. This work was also partially supported by FAPERJ's grant E-26/111.817/2012. 
LM acknowledge financial support from PRIN MIUR 2010-2011, project ``The Chemical and Dynamical Evolution of the Milky 
Way and Local Group Galaxies'', prot. 2010LY5N2T.

{}

\appendix
\section{Emission-line flux measurements}
In this section we present the observed emission-line fluxes and extinction corrected intensities 
measured in our sample of PNe. 

\begin{table}
\centering
\begin{minipage}{82mm}
{\tiny  
\caption{Observed fluxes and extinction corrected intensities of our PNe. Column
(1) gives the PN name; column (2) gives the observed \hb\ flux 
in units of 10$^{-16}$~erg cm$^{-2}$ s$^{-1}$; 
column (3) the nebular extinction
coefficient; columns (4) and (5) indicate the emitting ion and the rest frame wavelength
in \AA; columns (6), (7), and (8) give extinction corrected (I$_{\lambda}$) intensities, the 
relative error on the fluxes ($\Delta$F$_{\lambda}$) and the measured fluxes (F$_{\lambda}$). Both I$_{\lambda}$ and F$_{\lambda}$ are normalised to \hb=100. Upper limits on the line fluxes are marked with $:$.}
\begin{tabular}{@{}ccclcrrr@{}}
\hline
Id & F$_{{\rm H}\beta}$ & \cbeta\         & Ion & $\lambda$ (\AA) & I$_{\lambda}$ & $\Delta$F$_{\lambda}$  & F$_{\lambda}$ \\ 
   &                    & $\Delta$\cbeta\ &     &                 &               &        (\%)   	     &  	     \\ 
\hline
F1-8   & 13.58 & 0.458       & He~{\sc i}     &  3734  &   1.4  & 0.9 &   1.0	\\
       &       & $\pm$0.005  & He~{\sc i}     &  3872  &  61.8  & 1.7 &  47.7	\\
       &      &   	     & H~{\sc i}      &  3889  &  14.2  & 1.1 &  11.0	\\
       &      &   	     & [Ne~{\sc iii}] &  3968  &  40.1  & 1.5 &  31.8	\\
       &      &   	     & He~{\sc i}     &  4026  &   4.6  & 0.3 &   3.7	\\
       &      &   	     & H$\delta$      &  4100  &  27.0  & 0.9 &  22.1	\\
       &      &   	     & H$\gamma$      &  4340  &  59.2  & 1.1 &  51.7	\\
       &      &   	     & [O~{\sc iii}]  &  4363  &  32.7  & 1.0 &  28.7	\\
       &      &   	     & He~{\sc i}     &  4471  &   7.4  & 0.8 &   6.7	\\
       &      &      	     & He~{\sc ii}    &  4547  &   1.2  & 0.5 &   1.1	\\
       &      &   	     & He~{\sc ii}    &  4686  &  13.5  & 0.6 &  12.9	\\
       &      &   	     & Ar~{\sc iv}    &  4712  &   3.7  & 0.3 &   3.6	\\
       &      &      	     & Ar~{\sc iv}    &  4740  &   5.9  & 0.4 &   5.8	\\
       &      &   	     & H$\beta$       &  4861  & 100.0  & 0.8 & 100.0	\\
       &      &   	     & He~{\sc i}     &  4922  &   1.3  & 0.3 &   1.3	\\
       &      &   	     & [O~{\sc iii}]  &  4959  & 376.3  & 1.7 & 386.1	\\
       &      &   	     & [O~{\sc iii}]  &  5007  & 1120.7 & 2.7 & 1164.2  \\
       &      &   	     & He~{\sc i}     &  5016  &   3.5  & 0.3 &   3.6	\\
       &      &   	     & [Fe~{\sc iii}] &  5085  &   0.5  & 0.1 &   0.5	\\
       &      &   	     & [Fe~{\sc ii}]  &  5159  &   0.3  & 0.2 &   0.3	\\
       &      &   	     & [N~{\sc i}]    &  5200  &   0.7  & 0.2 &   0.7	\\
       &      &   	     & [Fe~{\sc iii}] &  5412  &   0.9  & 0.1 &   1.1	\\
       &      &   	     & C~{\sc iv}     &  5802  &   0.5  & 0.1 &   0.6	\\
       &      &   	     & [O~{\sc i}]    &  5577  &  11.4  & 0.3 &  13.5	\\
       &      &   	     & [Fe~{\sc ii}]  &  5582  &  18.1  & 0.2 &  21.6	\\
       &      &   	     & [N~{\sc ii}]   &  5755  &   1.8  & 0.2 &   2.2	\\
       &      &   	     & He~{\sc i}     &  5876  &  18.4  & 0.3 &  22.9	\\
       &      &   	     & [Si~{\sc ii}]  &  6347  &   0.4  & 0.2 &   0.5	\\
       &      &   	     & O~{\sc i}      &  6363  &   1.4  & 0.2 &   1.8	\\
       &      &   	     & [N~{\sc ii}]   &  6548  &  12.0  & 0.3 &  16.5	\\
       &      &   	     & H$\alpha$      &  6563  & 289.9  & 1.0 & 399.2	\\
       &      &   	     & [N~{\sc ii}]   &  6584  &  35.4  & 0.4 &  48.9	\\
       &      &   	     & He~{\sc i}     &  6678  &   4.5  & 0.3 &   6.3	\\
       &      &   	     & [S~{\sc ii}]   &  6717  &   1.5  & 0.2 &   2.2	\\
       &      &   	     & [S~{\sc ii}]   &  6731  &   2.7  & 0.2 &   3.8	\\
       &      &   	     & [Ar~{\sc v}]   &  7006  &   1.4  & 0.1 &   2.0	\\
       &      &   	     & He~{\sc i}     &  7065  &  13.7  & 0.1 &  20.3	\\
       &      &   	     & [Ar~{\sc iii}] &  7135  &  14.6  & 0.1 &  21.9	\\
       &      &   	     & [O~{\sc ii}]   &  7320  &   6.6  & 0.2 &  10.1	\\
       &      &   	     & [O~{\sc ii}]   &  7330  &   5.9  & 0.2 &   9.1	\\
       &      &   	     & [Ar~{\sc iii}] &  7751  &   3.0  & 0.1 &   4.9	\\
       &      &   	     & H~{\sc i}      &  8665  &   0.8  & 0.1 &   1.6	\\
       &      &   	     & H~{\sc i}       &  8750  &   1.4  & 0.2 &      2.7   \\
       &      &   	     & H~{\sc i}       &  8862  &   2.4  & 0.1 &      4.5   \\
       &      &   	     & H~{\sc i}       &  9015  &   1.4  & 0.2 &      2.7   \\
       &      &   	     & [S~{\sc iii}]   &  9069  &  13.3  & 0.3 &  26.5        \\
       &      &   	     & H~{\sc i}       &  9229  &   2.0  & 0.3 &   4.0   \\
       &      &   	     & He~{\sc i}      &  9526  &  46.9  & 0.3 & 100.1   \\
       &      &   	     & H~{\sc i}       &  9546  &   3.0  & 0.1 &   6.4   \\  	       
\hline      
F1-9 & 19.05 & 0.366       & He~{\sc i}     & 3587   & 14.6   & 0.9  & 11.2  \\  
     &      &  $\pm$0.003  & H~{\sc i}	    &  3667  &   3.6  &  0.7 &   2.8	\\
     &      &     	   & [O~{\sc ii}]   &  3726  &  18.2  &  1.2 &  14.4	\\
     &      &       	   & H~{\sc i}	    &  3771  &   2.1  &  0.8 &   1.7	\\
     &      &   	   & H~{\sc i}	    &  3798  &   2.5  &  0.9 &   2.0	\\
     &      &   	   & [Ne~{\sc iii}] &  3868  &  60.9  &  1.4 &  49.5	\\
     &      &   	   & H~{\sc i}      &  3889  &  11.4  &  1.0 &   9.3	\\
     &      &   	   & He~{\sc ii}    &  3922  &   2.3  &  1.0 &   1.9	\\
     &      &   	   & [Ne~{\sc iii}] &  3968  &  39.2  &  1.1 &  32.5	\\
     &      &      & [F~{\sc iv}]   &  4060  &   2.9  &  0.9 &   2.5	\\
     &      &      & [Si~{\sc ii}]  &  4069  &   4.8  &  0.9 &   4.1	\\
     &      &   	   & H$\delta$      &  4100  &  31.0  &  1.0 &  26.5	\\
     &      &      & H$\gamma$      &  4340  &  56.7  &  1.1 &  50.9	\\
     &      &   	   & [O~{\sc iii}]  &  4363  &  19.1  &  0.7 &  17.2	\\
     &      &   	   & O~{\sc ii}     &  4417  &   1.1  &  0.4 &   1.0	\\
     &      &   	   & He~{\sc i}     &  4471  &   6.0  &  0.8 &   5.5	\\
     &      &              & He~{\sc ii}    &  4686  &   5.5  &  0.8 &   5.3	\\
     &      &   	   & Ar~{\sc iv}    &  4712  &   2.9  &  0.3 &   2.8	\\
     &      &   	   & Ar~{\sc iv}    &  4740  &   3.3  &  0.3 &   3.2	\\
     &      &   	   & H$\beta$	    &  4861  & 100.0  &  0.7 &  100.0	\\
     &      &   	   & He~{\sc ii}    &  4922  &   1.2  &  0.5 &   1.2    \\
     &      &   	   & [O~{\sc iii}]  &  4959  & 404.5  &  1.5 &  412.9	\\
     &      &   	   & [O~{\sc iii}]  &  5007  & 1207.4 &  2.6 & 1244.6  \\
     &      &   	   & [N~{\sc ii}]   &  5755  &   0.8  &  0.4 &   1.0	\\
     &      &   	   & He~{\sc i}     &  5876  &  10.0  &  1.6 &  11.9	\\
     &      &   	   & [K~{\sc iv}]   &  6102  &   0.2  &  0.1 &   0.2	\\
     &      &   	   & He~{\sc ii}    &  6171  &   0.3  &  0.1 &   0.4	\\
     &      &   	   & [O~{\sc i}]    &  6300  &   4.8  &  0.1 &   6.0	\\
     &      &   	   & [S~{\sc iii}]  &  6312  &   2.7  &  0.1 &   3.4	\\
     &      &   	   & O~{\sc i}      &  6363  &   1.5  &  0.1 &   1.9	\\
     &      &   	   & [Si~{\sc ii}]  &  6371  &   0.2  &  0.1 &   0.3	\\
     &      &   	   & [N~{\sc ii}]   &  6548  &  13.5  &  0.1 &  17.5	\\
     &      &   	   & H$\alpha$      &  6563  & 289.3  &  0.2 & 373.6	\\
     &      &   	   & [N~{\sc ii}]   &  6584  &  40.7  &  0.5 &  52.6	\\
\end{tabular}
}
\end{minipage}
\label{tabPN_flux}
\end{table}
\begin{table}
\centering
\begin{minipage}{82mm}
{\tiny 
\contcaption{}
\begin{tabular}{@{}llllrrrr@{}}
\hline
Id & F$_{{\rm H}\beta}$ & \cbeta\ & Ion & $\lambda$ (\AA) & I$_{\lambda}$ & $\Delta$F$_{\lambda}$ & F$_{\lambda}$ \\ 
   &                    & $\Delta$\cbeta\ &     &                 &               &        (\%)   	     &  \\
\hline   	    
     &      &   	   & He~{\sc i}     &  6678  &   3.8  &  0.1 &   5.0	\\
     &      &   	   & [S~{\sc ii}]   &  6717  &   3.0  &  0.1 &   4.0	\\
     &      &   	   & [S~{\sc ii}]   &  6731  &   5.0  &  0.1 &   6.6	\\
     &      &   	   & He~{\sc i}     &  7065  &   9.7  &  0.1 &  13.3	\\
     &      &   	   & He~{\sc i}     &  7281  &   1.1  &  0.1 &   1.6	\\
     &      &   	   & [O~{\sc ii}]   &  7320  &   6.4  &  0.1 &   9.0	\\
     &      &   	   & [O~{\sc ii}]   &  7330  &   5.4  &  0.1 &   7.7	\\
     &      &   	   & [Ar~{\sc iii}] &  7751  &   3.3  &  0.2 &   4.9	\\
     &      &   	   & H~{\sc i}      &  8467  &   0.5  &  0.1 &   0.8	\\
     &      &   	   & [Cl~{\sc iii}] &  8500  &   0.6  &  0.1 &   0.9	\\
     &      &   	   & H~{\sc i}	    &  8546  &   0.9  &  0.2 &   1.4	\\
     &      &   	   & [Cl~{\sc ii}]  &  8579  &   0.2  &  0.1 &   0.3	\\
     &      &   	   & H~{\sc i}      &  859  &   0.7  &  0.2 &   1.2	\\
     &      &   	   & H~{\sc i}      &  8665  &   1.4  &  0.2 &   2.3	\\
     &      &   	   & N~{\sc i}      &  8680  &   0.7  &  0.2 &   1.1	\\
     &      &   	   & H~{\sc i}      &  8750  &   1.1  &  0.2 &   1.9	\\
     &      &   	   & H~{\sc i}      &  8862  &   1.3  &  0.2 &   2.2	\\
     &      &   	   & H~{\sc i}      &  9015  &   1.5  &  0.2 &   2.6	\\
     &      &   	   & [S~{\sc iii}]  &  9069  &  12.7  &  0.2 &  22.0	\\
     &      &   	   & He~{\sc i}     &  9210  &   0.3  &  0.1 &   0.6	\\
     &      &   	   & H~{\sc i}      &  9229  &   2.0  &  0.2 &   3.6	\\
     &      &   	   & He~{\sc i}     &  9526  &  32.4  &  0.5 &  59.3	\\
     &      &   	   & H~{\sc i}      &  9546  &   2.5  &  0.3 &   4.6	\\
     &      &              & H~{\sc i}      &  10049 &   2.5  &  0.1 &   4.8	\\ 
\hline
F1-10& 1.715& 0.306        & He~{\sc i}     &  3734  &  32.2 & 8.8 &  26.4  \\
     &      & $\pm$0.022   & H~{\sc i}      &  3889  &  49.9 &  8.0 &  42.2  \\
     &      &   	   & [Ne~{\sc iii}] &  3968  &  25.3 &  8.3 &  21.7  \\
     &      &   	   & H$\gamma$      &  4340  &  59.4 &  8.8 &  54.3  \\
     &      &   	   &  O~{\sc ii}    &  4349  &  22.2 &  9.7 &  20.3  \\
     &      &   	   & [O~{\sc iii}]  &  4363  &   9.8 &  8.3 &  8.9  \\
     &      &   	   & [Fe~{\sc ii}]		    &  4413  &  19.1 & 14.1 &  17.6  \\
     &      &   & H$\beta$       &  4861  & 100.0 & 4.0 & 100.0  \\
     &      &   & [O~{\sc iii}]  &  4959  & 246.7 & 4.1 & 251.0  \\
     &      &   & [O~{\sc iii}]  &  5007  & 749.0 & 5.7 & 768.5  \\
     &      &   & [N~{\sc i}]		 &  5198  &   4.9 & 2.3 &   5.2  \\
     &      &   & [Fe~{\sc ii}]		 &  5477  &   2.3 & 2.6 &   2.5  \\
     &      &   &  [Fe~{\sc ii}]              &  5496  &   2.0 & 2.5 &   2.2  \\
     &      &   & [Cl~{\sc iii}] &  5518  &   1.6 & 2.9 &   1.8  \\
     &      &   & [Fe~{\sc ii}]  &  5582  & 460.3 & 5.7 & 517.7  \\      
     &      &   & He~{\sc i}     &  5876  &  11.0 & 1.1 &  12.8  \\
     &      &   & [O~{\sc i}]    &  6300  &   4.7 & 1.1 &   5.7  \\
     &      &   & [N~{\sc ii}]   &  6548  &  73.5 & 4.0 &  90.8  \\
     &      &   &  H$\alpha$	   &  6563  & 288.9 & 4.0 & 357.8  \\
     &      &   & He~{\sc i}     &  7065  &   5.4 & 1.5 &   7.1  \\
     &      &   & [Ar~{\sc iii}] &  7135  &   9.1 & 1.3 &  11.9  \\
     &      &   & [Fe~{\sc ii}]		 &  7220  &   7.7 & 1.3 &  10.2  \\
     &      &   & [O~{\sc ii}]   &  7330  &  11.9 & 1.3 &  15.9  \\
\hline
F1-13& 4.359 & 0.423  &  H~{\sc i} & 3657  &   3.8 & 11.1 &   2.8  \\
    &       &  $\pm$0.010  & [O~{\sc ii}]   &  3729  &   11.2 & 3.8 &	8.5  \\
     &      &   	   & H11	    &  3771  &   10.6 & 4.9 &	8.2  \\
     &      &   	   & He~{\sc i}     &  3806  &   11.9 & 3.8 &	9.2  \\
     &      &   	   & [Ne~{\sc iii}] &  3868  &  25.2 & 2.8 &  19.8  \\
     &      &   	   & H$\gamma$      &  4340  &  64.2 & 3.8 &  56.6  \\
     &      &   	   &  O~{\sc ii}    &  4349  &   8.4 & 3.5 &   7.4  \\
     &      &   	   & [O~{\sc iii}] &  4363  &  10.0 & 3.5 &   8.8  \\
     &      &   	   & H$\beta$	    &  4861  & 100.0 & 1.9 & 100.0  \\
     &      &   	   & [O~{\sc iii}]  &  4959  & 216.6 & 2.7 & 221.7  \\
     &      &   	   & [O~{\sc iii}]  &  5007  & 643.8 & 3.8 & 666.7  \\
     &      &      & He~{\sc i}     &  5876  &   9.7 & 3.1 &  11.9  \\
     &      &   	   & [N~{\sc ii}]   &  6548  &   2.1 & 0.5 &   2.8  \\
     &      &   	   & H$\alpha$      &  6563  & 289.7 & 1.1 & 389.3  \\
     &      &   	   & [N~{\sc ii}]   &  6584  &   6.3 & 0.3 &   8.5  \\
     &      &   	   & He~{\sc i}     &  6678  &   3.5 & 0.4 &   4.8  \\
     &      &   	   & He~{\sc i}     &  7065  &   6.2 & 0.5 &   9.0  \\
     &      &   	   & [Ar~{\sc iii}] &  7135  &   9.3 & 0.4 &  13.5  \\
     &      &   	   & [O~{\sc ii}]   &  7320  &   1.3 & 0.6 &   2.0  \\
     &      &   	   & [O~{\sc ii}]   &  7330  &   1.7 & 0.5 &   2.6  \\
     &      &   	   & Si~{\sc i}     &  7538  &   4.2 & 0.4 &   6.5  \\
     &      &   	   & [Ar~{\sc iii}] &  7751  &   1.8 & 0.7 &   2.9  \\
     &      &   	   & [Cr~{\sc ii}]  &  8000  &   1.4 & 0.7 &   2.3  \\
     &      &   	   & N~{\sc i}      &  8703  &   2.6 & 0.3 &   4.6  \\
     &      &   	   & N~{\sc i}      &  8712  &   7.8 & 0.5 &  14.0  \\
     &      &   	   & H~{\sc i}      &  9229  &   2.3 & 0.4 &   4.5  \\
     &      &   	   & [S~{\sc iii}]  &  9069  &   7.8 & 0.5 &  14.8  \\
     &      &   	   & He~{\sc i}     &  9526  &  19.3 & 0.7 &  38.8  \\
     &      &   	   & H~{\sc i}      &  9546  &   2.1 & 0.4 &   4.2  \\
     &      &   	   & [C~{\sc i}]    &  9850  &   3.9 & 0.7 &   8.3  \\
\hline
F1-14 & 15.01 & 0.403       & [O~{\sc ii}]   &  3729  &  17.2    &  2.1  &  13.2 \\
     &       &  $\pm$0.004  & He~{\sc i}     &  3734  &  12.3	 &  1.3  &   9.5  \\
     &     &    	    &	He~{\sc i}	     &  3785  &   1.4	 &  1.5  &   1.1 \\
     &     &    	    & H~{\sc i}	     &  3798  &   2.6	 &  1.9  &   2.0 \\
     &     &    	    & He~{\sc i}     &  3872  &  37.6	 &  1.2  &  29.9 \\
     &     &    	    & H~{\sc i}      &  3889  &  11.1	 &  0.7  &   8.9 \\
     &     &    	    & [Ne~{\sc iii}] &  3968  &  29.4	 &  1.2  &  23.9 \\
     &     &    	    &	He~{\sc i}     &  4009 &   5.3	 &  0.8  &   4.4 \\
     &     &    	    & H$\delta$      &  4100  &  32.5	 &  1.1  &  27.3 \\
     &     &    	    & H$\gamma$      &  4340  &  61.6	 &  1.3  &  54.7 \\
     &     &    	    & [O~{\sc iii}]  &  4363  &   10.7    &  0.7  &  9.5 \\
     &     &    	    & He~{\sc i}     &  4471  &   6.3	 &  0.7  &   5.7 \\
     &     &    	    &	[Fe~{\sc ii}]	&  4874  &   1.5	 &  0.3  &   1.5 \\
     &     &    	    & H$\beta$       &  4861  & 100.0	 &  0.8  & 100.0 \\ 
     &     &    	    &	He~{\sc ii}     &  4922  &   1.8	 &  0.3  &   1.8 \\ 
     &     &    	    & [O~{\sc iii}]  &  4959  & 300.0	 &  1.4  & 306.9 \\ 
     &     &    	    & [O~{\sc iii}]  &  5007  & 889.7	 &  2.5  & 920.1 \\
\end{tabular}
}
\end{minipage}
\label{tabPN_flux}
\end{table}
\begin{table}
\centering
\begin{minipage}{82mm}
{\tiny
\contcaption{}
\begin{tabular}{@{}ccclcrrr@{}}
\hline
Id & F$_{{\rm H}\beta}$ & \cbeta\ & Ion & $\lambda$ (\AA) & I$_{\lambda}$ & $\Delta$F$_{\lambda}$ & F$_{\lambda}$ \\ 
   &                    & $\Delta$\cbeta\ &     &                 &               &        (\%)   	     &  \\	     \hline 
     &     &    	    & He~{\sc i}     &  5016  &   2.7	 &  0.3  &   2.8  \\
     &     &    	    & [Fe~{\sc ii}]  &  5582  &  33.3	 &  0.6  &  38.8 \\
     &     &    	    & [N~{\sc ii}]   &  5755  &   0.2	 &  0.2  &   0.2  \\
     &     &    	    & He~{\sc i}     &  5876  &   9.8	 &  0.4  &  11.9  \\
     &     &    	    & [O~{\sc i}]    &  6300  &   1.3	 &  0.1  &   1.7  \\
     &     &    	    & [S~{\sc iii}]  &  6312  &   0.6	 &  0.0  &   0.8  \\
     &     &    	    & [O~{\sc i}]    &  6363  &   0.6	 &  0.1  &   0.7 \\
     &     &    	    & [N~{\sc ii}]   &  6548  &   6.1	 &  0.1  &   8.1 \\
     &     &    	    & H$\alpha$      &  6563  & 289.5	 &  0.4  & 383.7 \\
     &     &    	    & [N~{\sc ii}]   &  6584  &  18.7	 &  0.2  &  24.9 \\
     &     &   & He~{\sc i}     &  6678  &   4.4    &  0.2  &	5.9 \\
     &     &   & [S~{\sc ii}]   &  6717  &   2.0    &  0.2  &	2.7 \\
     &     &   & [S~{\sc ii}]   &  6731  &   3.5    &  0.2  &	4.8 \\
     &     &   & [Fe~{\sc ii}]	&  6946  &   1.4    &  0.1  &	1.9 \\
     &     &   & He~{\sc i}     &  7065  &   8.5    &  0.2  &  12.0 \\   
     &     &   & [Ar~{\sc iii}] &  7135  &  12.3    &  0.2  &  17.5  \\  
     &     &   & He~{\sc i}     &  7281  &   1.3    &  0.2  &	1.9  \\ 
     &     &   & [O~{\sc ii}]   &  7320  &   3.5    &  0.1  &	5.1 \\
     &     &   & [O~{\sc ii}]   &  7330  &   2.8    &  0.2  &	4.0  \\
     &     &   & [Ar~{\sc iii}] &  7751  &   2.5    &  0.2  &	3.9 \\
     &     &   &H~{\sc i}	&  8598  &   0.8    &  0.2  &	1.3 \\
     &     &   & H~{\sc i}      &  8665  &   1.3    &  0.2  &	2.2  \\
     &     &   & H~{\sc i}      &  8750  &   0.8    &  0.2  &	1.5  \\
     &     &   & H~{\sc i}      &  8862  &   1.3    &  0.2  &	2.4  \\
     &     &   & H~{\sc i}      &  9015  &   1.5    &  0.1  &	2.7 \\
     &     &   & [S~{\sc iii}]  &  9069  &  14.1    &  0.2  &  25.8 \\
     &     &   & H~{\sc i}      &  9229  &   1.9    &  0.2  &	3.6 \\
     &     &   & [S~{\sc iii}]  &  9530  &  32.7    &  0.4  &  63.7 \\
     &     &   & H~{\sc i}      &  9546  &   2.2    &  0.3  &	4.3 \\
     &     &   & He~{\sc i}     & 10031  &   1.6    &  0.1  &	3.2  \\ 
     &     &   & H~{\sc i}      & 10049  &   3.2    &  0.2  &	6.6  \\ 
\hline
F1-15 & 12.75 & 0.448        & [O~{\sc ii}]   &  3726  &  28.9  & 1.3 & 21.6\\
    &       &  $ \pm$ 0.005  & [Ne~{\sc iii}] &  3868  &   4.6 & 1.1 &   3.6\\  
     &     &       	     &  H~{\sc i}    &  3889  &   7.5 & 1.2 &   5.8   \\
     &     &        	     & [Ne~{\sc iii}] &  3968  &  13.4 & 1.3 &  10.6   \\
     &     &        	     & He~{\sc i}     &  4009  &   3.1 & 1.0 &   2.5   \\
     &     &        	     & H$\delta$      &  4100  &  27.9 & 1.3 &  23.0   \\
     &     &        	     & O~{\sc ii}     &  4157  &   3.7 & 0.7 &   3.1   \\
     &     &        	     & H$\gamma$      &  4340  &  56.6 & 1.2 &  49.6   \\
     &     &        	     & [O~{\sc iii}]  &  4363  &   2.5 & 0.6 &   2.2   \\
     &     &        	     & He~{\sc i}     &  4471  &   5.5 & 0.6 &   5.0   \\
     &     &        	     & H$\beta$       &  4861  & 100.0 & 1.0 & 100.0   \\
     &     &        	     & [O~{\sc iii}]  &  4959  & 105.2 & 1.0 & 107.9   \\
     &     &        	     & [O~{\sc iii}]  &  5007  & 311.5 & 1.7 & 323.3   \\
     &     &        	     & He~{\sc i}     &  5016  &   2.1 & 0.4 &   2.1   \\
     &     &   & [N~{\sc ii}]   &  5755  &   1.4 & 0.1 &   1.7   \\	
     &     &   & [O~{\sc i}]    &  6300  &   1.1 & 0.1 &   1.5   \\	
     &     &   & [S~{\sc iii}]  &  6312  &   0.6 & 0.1 &   0.8    \\	
     &     &   & [Si~{\sc ii}]  &  6347  &   0.3 & 0.2 &   0.3   \\	
     &     &   & [O~{\sc i}]    &  6363  &   0.3 & 0.1 &   0.4    \\	
     &     &   & [N~{\sc ii}]   &  6548  &  26.8 & 0.3 &  36.6    \\	
     &     &   & H$\alpha$      &  6563  & 289.8 & 0.8 & 396.4    \\	
     &     &   & [N~{\sc ii}]   &  6584  &  80.5 & 0.5 & 110.5    \\	
     &     &   & He~{\sc i}     &  6678  &   3.9 & 0.2 &   5.4   \\	
     &     &   & [S~{\sc ii}]   &  6717  &   1.6 & 0.3 &   2.2   \\	
     &     &   & [S~{\sc ii}]   &  6731  &   3.0 & 0.3 &   4.2   \\	
     &     &   & He~{\sc i}     &  7065  &   8.9 & 0.2 &  13.1   \\	
     &     &   & [Ar~{\sc iii}] &  7135  &  11.4 & 0.3 &  16.9   \\	
     &     &   & He~{\sc i}     &  7281  &   1.0 & 0.2 &   1.6   \\	
     &     &   & [O~{\sc ii}]   &  7320  &  16.9 & 0.3 &  25.7   \\	
     &     &   & [O~{\sc ii}]   &  7330  &  14.3 & 0.3 &  21.9   \\	
     &     &   & [Ar~{\sc iii}] &  7751  &   2.6 & 0.3 &   4.2   \\	
     &     &   & H~{\sc i}      &  8272  &   0.3 & 0.2 &   0.5    \\	
     &     &   & O~{\sc i}      &  8447  &   1.8 & 0.2 &   3.2    \\	
     &     &   & H~{\sc i}      &  8665  &   1.4 & 0.2 &   2.6   \\   
     &     &   & H~{\sc i}      &  8750  &   2.1 & 0.2 &   3.9   \\   
     &     &   & H~{\sc i}      &  8862  &   1.7 & 0.3 &   3.3   \\   
     &     &   & H~{\sc i}      &  9015  &   1.7 & 0.2 &   3.4   \\   
     &     &   & [S~{\sc iii}]  &  9069  &  16.8 & 0.3 &  33.0   \\   
     &     &   & H~{\sc i}      &  9229  &   2.5 & 0.3 &   4.9    \\  
     &     &   &He~{\sc i}	&  9526  &  41.5 & 0.4 &  87.1    \\  
     &     &   & [C~{\sc i}]	&  9850  &   0.9 & 0.2 &   2.0   \\   
     &     &   & H~{\sc i}      & 10049  &   2.7 & 0.2 &   6.1   \\  
\hline
F1-16 & 16.54 & 0.409  & He~{\sc i}     &  3806  &  4.0  & 0.9  & 3.1 \\
    &       &  $\pm$0.005   & H~{\sc i}	     &  3835  &    4.9 &  0.5 &   3.9 \\
     &     &       	    & [Si~{\sc ii}]  &  3863  &    3.6 &  0.6 &   2.9 \\
     &     &       	    & [Ne~{\sc iii}] &  3868  &   42.3 &  1.3 &  33.6 \\
     &     &       	    & [Ne~{\sc iii}] &  3968  &   32.4 &  1.1 &  26.3 \\
     &     &       	    & H$\delta$      &  4100  &   29.4 &  1.3 &  24.6 \\
     &     &       	    & C~{\sc ii}     &  4270  &    2.9 &  0.6 &   2.5 \\
     &     &       	    & H$\gamma$      &  4340  &   59.2 &  1.1 &  52.5 \\
     &     &       	    & [O~{\sc iii}]  &  4363  &   10.8 &  0.7 &   9.6 \\
     &     &       	    & He~{\sc i}     &  4471  &    6.5 &  1.0 &   6.0 \\     	
     &     &   & Ar~{\sc iv}    &  4711  &    1.5 &  0.8 &   1.4 \\		      
     &     &   & Ar~{\sc iv}    &  4740  &    1.6 &  0.7 &   1.6 \\		      
     &     &   &	 S~{\sc ii}	&  4792  &    1.9 &  0.5 &   1.9 \\		      
     &     &   &	He~{\sc i}	&  4922  &    1.5 &  0.7 &   1.5 \\		      
     &     &   & H$\beta$       &  4861  &  100.0 &  0.9 & 100.0  \\		      
     &     &   &	He~{\sc i}	&  4922  &    1.5 &  0.7 &   1.5 \\		      
     &     &   & [O~{\sc iii}]  &  4959  &  286.0 &  1.6 & 292.6 \\		      
     &     &   & [O~{\sc iii}]  &  5007  &  850.3 &  2.7 & 879.7  \\		      
     &     &   & He~{\sc i}     &  5016  &    2.5 &  0.9 &   2.6  \\		      
     &     &   & [Fe~{\sc ii}]  &  5582  &   29.8 &  0.3 &  34.8  \\		      
     &     &   & [N~{\sc ii}]   &  5755  &    0.7 &  0.1 &   0.8 \\
     &     &   & He~{\sc i}     &  5876  &   10.3 &  0.2 &  12.6  \\		      		      
     &     &   & [O~{\sc i}]    &  6300  &    2.9 &  0.1 &   3.7 \\		      
\end{tabular}
}
\end{minipage}
\label{tabPN_flux}
\end{table}
\begin{table}
\centering
\begin{minipage}{82mm}
{\tiny 
\contcaption{}
\begin{tabular}{@{}ccclcrrr@{}}
\hline
Id & F$_{{\rm H}\beta}$ & \cbeta\ & Ion & $\lambda$ (\AA) & I$_{\lambda}$ & $\Delta$F$_{\lambda}$ & F$_{\lambda}$ \\ 
   &                    & $\Delta$\cbeta\ &     &                 &               &        (\%)   	     &  \\	    
\hline 
     &     &   & [S~{\sc iii}]  &  6312  &    1.0 &  0.1 &   1.2 \\		      
     &     &   & [O~{\sc i}]    &  6363  &    1.1 &  0.1 &   1.4 \\		      
     &     &   & He~{\sc ii}    &  6406  &    0.2 &  0.1 &   0.3 \\		      
     &     &   &	 C~{\sc ii}	&  6454  &    0.1 &  0.1 &   0.1 \\		      
     &     &   & [N~{\sc ii}]   &  6548  &   11.8 &  0.1 &  15.7 \\		      
     &     &   & H$\alpha$      &  6563  &  289.5 &  0.5 & 385.3 \\		      
     &     &   & [N~{\sc ii}]   &  6584  &   35.2 &  0.2 &  46.9 \\		      
     &     &   & He~{\sc i}     &  6678  &    4.0 &  0.1 &   5.4  \\		      
     &     &   & [S~{\sc ii}]   &  6717  &    1.9 &  0.1 &   2.6  \\		      
     &     &   & [S~{\sc ii}]   &  6731  &    3.4 &  0.1 &   4.6  \\		      
     &     &   & He~{\sc i}     &  7065  &   11.7 &  0.1 &  16.6   \\		      
     &     &   & [Ar~{\sc iii}] &  7135  &   15.2 &  0.2 &  21.9  \\		      
     &     &   & He~{\sc i}     &  7281  &    1.0 &  0.1 &   1.4  \\		      
     &     &   & [O~{\sc ii}]   &  7320  &    7.7 &  0.1 &  11.2   \\		      
     &     &   & [O~{\sc ii}]   &  7330  &    6.2 &  0.2 &   9.2   \\		      
     &     &   & [Ni~{\sc ii}]  &  7378  &    0.6 &  0.2 &   0.9   \\		      
     &     &   & [Ar~{\sc iii}] &  7751  &    3.3 &  0.1 &   5.1  \\		  
     &     &   & H~{\sc i}      &   8467  &   1.8 & 0.1 &   3.0  \\		     
     &     &   &	H~{\sc i}	&   8545  &   0.6 & 0.1 &   1.0  \\		     
     &     &   &	H~{\sc i}	&   8598  &   0.8 & 0.1 &   1.4  \\		     
     &     &   & [C~{\sc i}]    &   8727  &   0.5 & 0.1 &   0.9  \\		     
     &     &   & H~{\sc i}      &   8750  &   1.1 & 0.1 &   2.0  \\		     
     &     &   & H~{\sc i}      &   8862  &   1.6 & 0.2 &   3.0  \\		     
     &     &   & H~{\sc i}      &   9015  &   2.0 & 0.2 &   3.6   \\		     
     &     &   & [S~{\sc iii}]  &   9069  &  17.6 & 0.2 &  32.6   \\		     
     &     &   & H~{\sc i}      &   9229  &   2.7 & 0.2 &   5.1  \\		      
     &     &   &	He~{\sc i}	&   9526  &  31.9 & 0.5 &  62.8  \\		      
     &     &   & H~{\sc i}      &   9546  &   2.5 & 0.2 &   4.8  \\		      
     &     &   & H~{\sc i}      &  10049  &   3.0 & 0.3 &   6.3  \\	 
\hline
F1-20 & 13.37 & 0.531  &  He~{\sc i}&   3478 & 5.1 & 0.1 & 3.4 \\
    &       &  $\pm$0.005   & [O~{\sc ii}]   &  3726  &   9.6 & 0.5 &   6.8  \\     	     
     &     &        	   & [Ne~{\sc iii}] &  3868  &  32.3 & 3.4 &  23.9  \\      
     &     &        	   & H~{\sc i}    &  3889  &   9.8 & 1.1 &   7.3  \\
     &     &    	   & [Ne~{\sc iii}] &  3968  &  15.4 & 3.1 &  11.7  \\      
     &     &        	   & He~{\sc i}     &  4026  &   5.3 & 0.1 &   4.1   \\     
     &     &        	   & H$\delta$      &  4100  &  28.5 & 3.5 &  22.6  \\      
     &     &        	   &	[Fe~{\sc ii}]	    &  4177  &  15.0 & 1.5 &  12.2  \\      
     &     &        	   & H$\gamma$      &  4340  &  54.3 & 2.2 &  46.4   \\     
     &     &        	   & [O~{\sc iii}]  &  4363  &   4.5 & 1.2 &   3.8   \\     
     &     &        	   & He~{\sc i}     &  4471  &   5.7 & 0.7 &   5.1   \\     
     &     &        	   & H$\beta$	    &  4861  & 100.0 & 0.9 & 100.0  \\      
     &     &        	   &		He~{\sc i}    &  4922  &   1.3 & 0.1 &   1.3  \\      
     &     &        	   & [O~{\sc iii}]  &  4959  & 269.7 & 1.5 & 277.7  \\      
     &     &        	   & [O~{\sc iii}]  &  5007  & 800.4 & 2.5 & 836.2  \\      
     &     &        	   & He~{\sc i}     &  5016  &   2.9 & 0.1 &   3.1  \\      
     &     &        	   & [N~{\sc i}]    &  5200  &   0.6 & 0.1 &   0.7  \\      
     &     &        	   & C~{\sc iv}     &  5802  &   0.7 & 0.2 &   0.9  \\      
     &     &        	   & He~{\sc i}     &  5876  &  14.6 & 0.2 &  18.9  \\      
     &     &        	   & [O~{\sc i}]    &  6300  &   2.8 & 0.2 &   3.8   \\     
     &     &        	   & [S~{\sc iii}]  &  6312  &   1.5 & 0.2 &   2.1   \\     
     &     &        	   & [O~{\sc i}]    &  6363  &   1.1 & 0.2 &   1.5   \\     
     &     &        	   & [N~{\sc ii}]   &  6548  &  10.4 & 0.2 &  15.0    \\    
     &     &        	   & H$\alpha$      &  6563  & 290.4 & 0.8 & 420.7   \\     
     &     &        	   & [N~{\sc ii}]   &  6584  &  32.6 & 0.3 &  47.4   \\     
     &     &        	   & He~{\sc i}     &  6678  &   4.0 & 0.2 &   6.0    \\    
     &     &        	   & [S~{\sc ii}]   &  6717  &   1.4 & 0.2 &   2.0    \\    
     &     &        	   & [S~{\sc ii}]   &  6731  &   2.0 & 0.2 &   3.0    \\    
     &     &        	   & He~{\sc i}  &  7065  &  11.8 & 0.2 &  18.7   \\       
     &     &   & He~{\sc i}  &  7065  &  13.1 & 0.2 &  20.6  \\		 
     &     &   & [Ar~{\sc iii}] &  7135  &  15.3 & 0.3 &  24.4   \\		    
     &     &   &	He~{\sc i}+[Fe~{\sc ii}]	&  7281  &   1.4 & 0.2 &   2.3   \\		    
     &     &   & [O~{\sc ii}]   &  7320  &   8.6 & 0.2 &  14.2   \\		    
     &     &   & [O~{\sc ii}]   &  7330  &   7.0 & 0.2 &  11.5   \\		    
     &     &   & [Ar~{\sc iii}] &  7751  &   3.4 & 0.2 &   6.0   \\		    
     &     &   &	N~{\sc i} 	&  8184  &   0.2 & 0.1 &   0.3    \\		    
     &     &   &	N~{\sc i} 	&  8193  &   0.2 & 0.1 &   0.4    \\		    
     &     &   &	H~{\sc i}	&  8359  &   0.3 & 0.2 &   0.7   \\		    
     &     &   &	H~{\sc i}	&  8545  &   1.1 & 0.2 &   2.3   \\		    
     &     &   &	H~{\sc i}	&  8595  &   0.5 & 0.2 &   1.0   \\		    
     &     &   & H~{\sc i}      &  8665  &   1.4 & 0.2 &   2.8   \\		    
     &     &   & H~{\sc i}      &  8750  &   1.1 & 0.2 &   2.3   \\		    
     &     &   &	He~{\sc i}+[Fe~{\sc ii}]	&  8830  &   5.2 & 0.2 &  11.2   \\		    
     &     &   &	H~{\sc i}	&  8862  &   1.3 & 0.2 &   2.8   \\		    
     &     &   &	He~{\sc i}	&  8966  &   2.9 & 0.2 &   6.3  \\		    
\hline 
F1-21 & 9.045 & 0.465      & [O~{\sc ii}]   &  3726  &  13.4 & 2.5 &   9.9  \\
    &       &  $\pm$0.006  & He~{\sc i}     &  3872  &  23.8 & 2.5 &  18.3  \\
     &     &         	   & H~{\sc i}      &  3889  &  10.8 & 2.0 &   8.3  \\
     &     &         	   & [Ne~{\sc iii}] &  3968  &  10.5 & 2.5 &   8.3  \\
     &     &         	   & H$\delta$      &  4100  &  28.7 & 1.8 &  23.5  \\
     &     &         	   & H$\gamma$      &  4340  &  65.4 & 0.3 &  57.0  \\
     &     &         	   & [O~{\sc iii}]  &  4363  &   7.2 & 1.6 &   6.3  \\
     &     &         	   &	 [Fe~{\sc ii}]+[O~{\sc ii}]	    &  4416  &   3.6 & 1.6 &   3.2  \\
     &     &         	   &	 O~{\sc ii}	    &  4673  &   3.9 & 1.8 &   3.7   \\
     &     &         	   & H$\beta$	    &  4861  & 100.0 & 1.3 & 100.0   \\
     &     &         	   &	 He~{\sc i}	    &  4922  &   1.8 & 0.7 &   1.9    \\
     &     &         	   & [O~{\sc iii}]  &  4959  & 185.1 & 1.3 & 189.9   \\
     &     &         	   & [O~{\sc iii}]  &  5007  & 542.5 & 2.5 & 564.0   \\
     &     &         	   & He~{\sc i}     &  5876  &  16.1 & 0.6 &  20.1    \\
     &     &         	   & [S~{\sc iii}]  &  6312  &   2.2 & 0.8 &   2.9    \\
     &     &         	   & [N~{\sc ii}]   &  6548  &   8.9 & 0.3 &  12.3    \\
     &     &         	   & H$\alpha$      &  6563  & 289.9 & 0.9 & 401.4    \\
     &     &         	   & [N~{\sc ii}]   &  6584  &  26.7 & 0.4 &  37.1   \\
     &     &         	   & He~{\sc i}     &  6678  &   4.6 & 0.3 &   6.4   \\
     &     &         	   & [S~{\sc ii}]   &  6717  &   1.3 & 0.3 &   1.8   \\
     &     &         	   & [S~{\sc ii}]   &  6731  &   2.3 & 0.4 &   3.2   \\
     &     &         	   & He~{\sc i}     &  7065  &   9.6 & 0.4 &  14.3   \\
\end{tabular}
}
\end{minipage}
\label{tabPN_flux}
\end{table}
\begin{table}
\centering
\begin{minipage}{82mm}
{\tiny 
\contcaption{}
\begin{tabular}{@{}ccclcrrr@{}}
\hline
Id & F$_{{\rm H}\beta}$ & \cbeta\ & Ion & $\lambda$ (\AA) & I$_{\lambda}$ & $\Delta$F$_{\lambda}$ & F$_{\lambda}$ \\ 
   &                    & $\Delta$\cbeta\ &     &                 &               &        (\%)   	     &  \\	     
\hline
     &     &         	   & [Ar~{\sc iii}] &  7135  &  10.1 & 0.3 &  15.2   \\
     &     &         	   & [O~{\sc ii}]   &  7320  &   8.6 & 0.4 &  13.4   \\
     &     &         	   & [O~{\sc ii}]   &  7330  &   7.2 & 0.3 &  11.1    \\
     &     &         	   & [Ar~{\sc iii}] &  7751  &   2.3 & 0.4 &   3.9    \\
     &     &         	   & [C~{\sc i}]    &  8727  &   0.8 & 0.4 &   1.5   \\
     &     &         	   & H~{\sc i}      &  8750  &   1.4 & 0.4 &   2.6   \\
     &     &         	   & H~{\sc i}      &  8862  &   1.4 & 0.3 &   2.8   \\
     &     &         	   & [S~{\sc iii}]  &  9069  &  15.7 & 1.2 &  31.7   \\
     &     &         	   & H~{\sc i}      &  9229  &   2.4 & 0.8 &   5.0   \\
     &     &         	   & [S~{\sc iii}] &  9530  &  32.6 & 1.0 &  70.5	\\
     &     &         	   & H~{\sc i}      &  9546  &   2.7 & 0.9 &   5.8   \\
\hline
F1-22 & 0.324 &  0.743  & [O~{\sc ii}]   &  3726  &  141.8  & 15.0 &  87.7  \\
   &       &  $\pm$0.079  & H$\beta$	   &  4861  &  100.0  & 15.9 & 100.0  \\
     &     &    	  & [O~{\sc iii}]  &  4959  &  374.5  & 20.6 & 390.4   \\ 
     &     &    	  & [O~{\sc iii}]  &  5007  & 1130.7  & 27.1 &  1202   \\ 
     &     &    	  & [Fe~{\sc ii}]  &  5582  & 2133.3  &  34. &  2836  \\
     &     &    	  & [N~{\sc ii}]   &  5755  &	 5.6  & 19.2 &   7.8  \\
     &     &              & [Fe~{\sc ii}] &  5781  &	11.6  & 16.4 &  16.2   \\
     &     &    	  & [O~{\sc i}] &  6300  &	55.2  &  6.5 &  87.2  \\
     &     &    	  & [O~{\sc i}] &  6363  &	18.2  &  6.5 &  29.2   \\ 
     &     &    	  & H$\alpha$	   &  6563  &  291.6  & 11.7 & 490.1  \\
     &     &    	  & [N~{\sc ii}]   &  6584  &	39.7  &  7.0 &  67.1  \\
     &     &    	  & He~{\sc i}     &  6678  &	12.2  &  7.5 &  21.0  \\
     &     &    	  & [S~{\sc ii}]   &  6717  &	18.7  & 10.7 &  32.6  \\
     &     &    	  & [S~{\sc ii}]   &  6731  &	16.3  & 14.0 &  28.6  \\  
     &     &    	  & He~{\sc i}     &  8777  & 38.5 & 7.5 & 109.8  \\	     
     &     &    	  & [S~{\sc iii}]  &  9069  & 11.4 & 9.3 &  34.8  \\	     
     &     &    	  & [S~{\sc iii}]  &  9530  & 24.9 & 8.9 &  84.9  \\	     
     &     &    	  & He~{\sc i}     & 10031  & 63.2 & 8.9 & 243.6  \\	          	  
\hline
F2-3 & 1.209 & 0.456  & H~{\sc i} &   3669 &  29.0 & 4.4 & 21.2  \\
    &       &  $\pm$0.025 &	He~{\sc i} & 3833  &  66.7 & 3.9 &  51.0  \\
     &     &    	  & [Si~{\sc ii}]  & 3863  &  36.7 & 3.9 &  28.3   \\
     &     &      	  & [Ne~{\sc iii}] & 3868  &  77.6 & 4.4 &  59.9   \\
     &     &      	  & O~{\sc ii}     & 4190  &  60.3 & 3.9 &  50.6   \\
     &     &      	  & H$\gamma$	   & 4340  &  90.0 & 4.5 &  78.6    \\
     &     &      	  & O~{\sc ii}     & 4676  &  42.0 & 3.9 &  40.0   \\
     &     &      	  & He~{\sc ii}    & 4686  &  44.2 & 4.5 &  42.2   \\
     &     &      	  & H$\beta$	   & 4861  & 100.0 & 5.1 & 100.0    \\
     &     &      	  & [O~{\sc iii}]  & 4959  & 191.1 & 4.6 & 196.0    \\
     &     &      	  & [O~{\sc iii}]  & 5007  & 579.1 & 5.0 & 601.5    \\
     &     &      	  & He~{\sc i}     & 5876  &  11.2 & 4.4 &  13.9    \\
     &     &      	  & [O~{\sc i}]    & 6300  &  10.5 & 3.8 &  13.8   \\
     &     &      	  & [O~{\sc i}]    & 6363  &   5.6 & 1.2 &   7.5   \\
     &     &      	  & [N~{\sc ii}]   & 6548  &  15.5 & 1.6 &  21.3   \\
     &     &      	  & H$\alpha$	   & 6563  & 289.8 & 2.5 & 398.7   \\
     &     &      	  & [N~{\sc ii}]   & 6584  &  48.9 & 4.3 &  67.4   \\
     &     &      	  & [S~{\sc ii}]   & 6717  &   2.5 & 1.9 &   3.6   \\
     &     &      	  & [S~{\sc ii}]   & 6731  &   3.7 & 1.4 &   5.2   \\
     &     &      	  & [K~{\sc iv}]   & 6792  &   3.2 & 1.5 &   4.6    \\
     &     &      	  & He~{\sc i}     & 7065  &   7.9 & 1.8 &  11.7    \\
     &     &      	  & [Ar~{\sc iii}] & 7135  &   6.8 & 1.8 &  10.1   \\
     &     &      	  & [O~{\sc ii}]   & 7320  &  10.6 & 2.4 &  16.3   \\
     &     &      	  & [O~{\sc ii}]   & 7330  &   9.2 & 1.9 &  14.1   \\
     &     &      	  & [S~{\sc iii}]  & 9530  &  22.2 & 1.9 &  47.3   \\
     &     &      	  & [S~{\sc iii}]  & 9069  &  11.3 & 1.8 &  22.4   \\    	
\hline
F2-4 & 7.479 & 0.120  & [O~{\sc ii}]   &  3729  &  61.4 & 1.7 & 56.8  \\
    &       &  $\pm$0.009   & He~{\sc i}     & 3872  &  11.6  & 0.9 &  10.8   \\     
     &     &  	     	    & H~{\sc i}      & 3889  &   4.9  & 0.6 &	4.6   \\     
     &     &  	     	    &He~{\sc i}    & 4009  &  31.0  & 1.1 &  29.3   \\     
     &     &  	     	    & H$\delta$      & 4100  &  36.8  & 1.3 &  35.0    \\    
     &     &  	     	    & H$\gamma$      & 4340  &  87.6  & 1.9 &  84.6   \\     
     &     &  	     	    & [O~{\sc iii}]  & 4363  &   1.3  & 0.5 &	1.3   \\     
     &     &  	     	    & [Fe~{\sc ii}] & 4457  &   6.4  & 0.8 &	6.3    \\    
     &     &  	     	    & He~{\sc i}     & 4471  &   9.6  & 0.9 &	9.4    \\    
     &     &  	     	    & H$\beta$       & 4861  & 100.0  & 1.9 & 100.0    \\    
     &     &  	     	    & [O~{\sc iii}]  & 4959  & 148.3  & 2.0 & 149.4    \\    
     &     &  	     	    & [O~{\sc iii}]  & 5007  & 316.9  & 3.0 & 320.1   \\     
     &     &  	     	    & [N~{\sc ii}]   & 5755  &   1.4  & 0.7 &	1.5   \\     
     &     &  	     	    & He~{\sc i}     & 5876  &  19.0  & 0.5 &  20.1   \\     
     &     &  	     	    & He~{\sc ii}    & 5932  &   4.7  & 0.3 &	5.0   \\     
     &     &  	     	    & [O~{\sc i}]    & 6300  &   2.3  & 0.2 &	2.5   \\     
     &     &  	     	    & [O~{\sc i}]    & 6363  &   0.9  & 0.5 &	1.0   \\     
     &     &  	     	    & [N~{\sc ii}]   & 6548  &  29.4  & 0.6 &  32.0   \\     
     &     &  	     	    & H$\alpha$      & 6563  & 287.7  & 0.9 & 313.0   \\     
     &     &  	     	    & [N~{\sc ii}]   & 6584  &  74.6  & 0.6 &  81.2   \\     
     &     &  	     	    & He~{\sc i}     & 6678??&   6.0  & 0.3 &	6.6    \\
     &     &  		    & He~{\sc i}     & 6678??&   5.1  & 0.3 &	5.6   \\     
     &     &  	     	    & [S~{\sc ii}]   & 6717  &   2.4  & 0.2 &	2.6    \\    
     &     &  	     	    & [S~{\sc ii}]   & 6731  &   3.7  & 0.3 &	4.0   \\     
     &     &  	     	    &[Fe~{\sc ii}]   & 6746  &   1.9  & 0.3 &	2.1   \\     
     &     &  	     	    & He~{\sc i}     & 7065  &  10.4  & 0.3 &  11.5   \\     
     &     &  	     	    & [Ar~{\sc iii}] & 7135  &  11.2  & 0.4 &  12.5   \\     
     &     &  	     	    & [O~{\sc ii}]   & 7320  &  19.8  & 0.4 &  22.2   \\ 
     &     &  		        & [O~{\sc ii}]   & 7330  &  12.6  & 0.4 &  14.2   \\ 
     &     &  		        & H~{\sc i}      & 8665  &   2.2  & 0.3 &	2.6   \\ 
     &     &  		        & H~{\sc i}      & 8750  &   1.9  & 0.3 &	2.3   \\
     &     &  		        & H~{\sc i}      & 8862  &   2.6  & 0.4 &	3.1   \\  
     &     &  		        & [S~{\sc iii}]  & 9069  &  25.7  & 0.4 &  30.8    \\
     &     &  		        & H~{\sc i}      & 9015  &   3.2  & 0.3 &	3.8    \\
     &     &  		        & [S~{\sc iii}]  & 9530  &  63.6  & 0.5 &  77.6   \\ 
     &     &  		        & H~{\sc i}      & 9229  &   4.0  & 0.4 &	4.8   \\ 
     &     &  		        & H~{\sc i}      & 9546  &   5.8  & 0.3 &	7.0   \\ 
\hline
\end{tabular}
}
\end{minipage}
\label{tabPN_flux}
\end{table}
\begin{table}
\centering
\begin{minipage}{82mm}
{\tiny 
\contcaption{}
\begin{tabular}{@{}ccclcrrr@{}}
\hline
Id & F$_{{\rm H}\beta}$ & \cbeta\ & Ion & $\lambda$ (\AA) & I$_{\lambda}$ & $\Delta$F$_{\lambda}$ & F$_{\lambda}$ \\ 
   &                    & $\Delta$\cbeta\ &     &                 &               &        (\%)   	     & \\ 	    
\hline
F2-7 & 0.761 & 0.000  & [Ne~{\sc iii}] &   3868  & 74.6 & 6.8 & 74.6  \\
   &       &  $\pm$0.068 & H~{\sc i}    &  3889  & 147.2 &  9.7  & 147.2	 \\       
     &     &       	 & [Ne~{\sc iii}] & 3968  &  72.2 & 10.9  &  72.2    \\ 	      
     &     &    	 & He~{\sc i}	  & 4120  & 436.7 & 14.1  & 436.7     \\	      
     &     &    	 & H$\gamma$	  & 4340  &  58.6 &  6.8  &  58.6    \\ 	     
     &     &   & [O~{\sc iii}]  & 4363  &   46.2   &  6.2  &   46.2   \\ 		
     &     &   & He~{\sc ii}    & 4686  &   57.8   &  6.8  &   57.8   \\ 		
     &     &   & H$\beta$       & 4861  &  100.0   &  9.3  &  100.0   \\ 		
     &     &   & [O~{\sc iii}]  & 4959  &  430.2   & 13.5  &  430.2   \\		 
     &     &   & [O~{\sc iii}]  & 5007  & 1350.9   & 19.9  & 1350.9   \\		 
     &     &   & [Fe~{\sc ii}]  & 5582  &  415.9   & 14.5  &  415.9   \\
     &     &   & [N~{\sc ii}]   & 6548  &    9.1   &  6.4  &	9.1   \\	 
     &     &   & H$\alpha$      & 6563  &  159.4   & 10.1  &  159.4   \\		 
     &     &   & [N~{\sc ii}]   & 6584  &   11.5   &  7.2  &   11.5   \\		 
\hline
F2-10 & 8.334 & 0.110  & [O~{\sc iii}]  &  3729 &  37.3 & 1.1 & 34.7  \\
    &       &  $\pm$0.008  & H~{\sc i}      &  3889  &   6.8  & 1.0 &	6.4  \\
     &     &     	   & He~{\sc i}     &  3806  &   3.4  & 1.0 &	3.1  \\
     &     &     	   & [Ne~{\sc iii}] &  3968  &   4.5  & 0.9 &	4.3  \\
     &     &     	   & H$\delta$      &  4100  &  18.6  & 1.2 &  17.8   \\
     &     &     	   & O~{\sc ii}     &  4157  &   4.5  & 0.9 &	4.3  \\
     &     &     	   & [O~{\sc iii}]  &  4363  &  61.1  & 1.6 &  59.1  \\
     &     &     	   &	 O~{\sc ii}  &  4366  &   3.5  & 0.8 &	3.4   \\
     &     &     	   & H$\beta$	    &  4861  & 100.0  & 1.7 & 100.0   \\
     &     &     	   & [O~{\sc iii}]  &  4959  &  77.3  & 1.5 &  77.8   \\
     &     &     	   & [O~{\sc iii}]  &  5007  & 214.7  & 2.0 & 216.7   \\
     &     &     	   & He~{\sc i}     &  5876  &   9.4  & 0.3 &	9.9  \\
     &     &     	   & [N~{\sc ii}]   &  6548  &  13.6  & 0.3 &  14.7  \\
     &     &     	   & H$\alpha$      &  6563  & 287.7  & 0.9 & 310.7  \\
     &     &     	   & [N~{\sc ii}]   &  6584  &  47.3  & 0.3 &  51.2  \\
     &     &     	   & He~{\sc ii}    &  6683  &   2.6  & 0.2 &	2.9  \\
     &     &     	   & [S~{\sc ii}]   &  6717  &   1.6  & 0.2 &	1.8  \\
     &     &     	   & [S~{\sc ii}]   &  6731  &   1.7  & 0.3 &	1.8  \\
     &     &     	   & O~{\sc ii}     &  6786  &   2.5  & 0.2 &	2.7  \\
     &     &     	   & [Ar~{\sc v}]   &  7006  &   0.8  & 0.2 &	0.9  \\
     &     &     	   & He~{\sc i}     &  7065  &   8.0  & 0.3 &	8.8   \\      
     &     &   & [Ar~{\sc iii}] &  7135  &   4.9  & 0.3 &   5.4  \\		   
     &     &   & He~{\sc i}     &  7281  &   1.4  & 0.2 &   1.6  \\		   
     &     &   & [O~{\sc ii}]   &  7320  &  16.5  & 0.5 &  18.3  \\		   
     &     &   & [O~{\sc ii}]   &  7330  &  12.2  & 0.4 &  13.6  \\		   
     &     &   & H~{\sc i}      &  8268  &   1.1  & 0.3 &   1.3  \\		   
     &     &   & [S~{\sc iii}]  &  9069  &  15.3  & 0.2 &  18.0  \\ 
\hline 
F2-11 & 2.456 & 0.000  & He~{\sc i} &  3926 &  23.1 & 1.2 & 23.1  \\
     &     & $\pm$0.014  & H$\delta$      &  4100  &  20.9  & 1.4 &  20.9  \\    	   
     &     &   & H$\gamma$      &  4340  &  46.9  & 1.7 &  46.9  \\		    
     &     &   & [O~{\sc iii}]  &  4363  &   5.9  & 0.9 &   5.9  \\		    
     &     &   & H$\beta$       &  4861  & 100.0  & 2.3 & 100.0  \\		    
     &     &   & [O~{\sc iii}]  &  4959  & 179.2  & 3.0 & 179.2  \\		    
     &     &   & [O~{\sc iii}]  &  5007  & 505.3  & 5.4 & 505.3   \\		    
     &     &   & He~{\sc i}     &  5876  &  10.6  & 1.1 &  10.6   \\		    
     &     &   & H$\alpha$      &  6563  & 260.4  & 2.4 & 260.4  \\		    
     &     &   & [N~{\sc ii}]   &  6584  &   3.0  & 0.8 &   3.0  \\		   
     &     &   & He~{\sc i}     &  6678  &   4.5  & 0.9 &   4.5  \\		    
     &     &   & [Ar~{\sc iii}] &  7135  &  10.7  & 1.4 &  10.7  \\		    
     &     &   & [S~{\sc iii}]  &  9069  &  19.6  & 1.8 &  19.6  \\		    
     &     &   & He~{\sc i}     &  9526  &  26.4  & 6.0 &  26.4  \\ 
\hline
F2-15 & 3.536 & 0.000  & He~{\sc i}     &  3188 &  14.4 & 1.5 & 14.4  \\
     &       &  $\pm$0.014   & H~{\sc i}      &  3734  &   5.7 & 1.1 &   5.7  \\
     &     &       	     & He~{\sc i}     &  3872  &  35.8 & 1.8 &  35.8  \\
     &     &       	     & H$\gamma$      &  4340  &  60.2 & 2.3 &  60.2  \\
     &     &       	     & [O~{\sc iii}]  &  4363  &   7.1 & 1.2 &   7.1  \\
     &     &       	     & N~{\sc ii}     &  4780  &  10.8 & 1.6 &  10.8  \\
     &     &       	     & H$\beta$       &  4861  & 100.0 & 2.9 & 100.0  \\
     &     &       	     & He~{\sc i}     &  4922  &   4.3 & 1.3 &   4.3   \\
     &     &       	     & He~{\sc i}     &  5048  &  13.1 & 1.8 &  13.1   \\
     &     &       	     & [Fe~{\sc iii}] &  5412  &   3.1 & 1.5 &   3.1  \\
     &     &       	     & [O~{\sc iii}]  &  4959  & 264.6 & 4.7 & 264.6  \\
     &     &       	     & [O~{\sc iii}]  &  5007  & 834.8 & 5.6 & 834.8  \\
     &     &       	     & He~{\sc i}     &  5876  &  11.3 & 0.7 &  11.3  \\
     &     &       	     & H$\alpha$      &  6563  & 273.8 & 1.1 & 273.8  \\
     &     &       	     & [N~{\sc ii}]   &  6584  &   7.0 & 0.8 &   7.0  \\ 
     &     &    	     & He~{\sc i}     &  6678  &   3.0 & 0.6 &   3.0  \\
     &     &    	     & He~{\sc i}     &  7065  &   4.9 & 0.6 &   4.9  \\
     &     &    	     & [Ar~{\sc iii}] &  7135  &   9.1 & 0.5 &   9.1  \\
     &     &    	     & H~{\sc i}      &  8750  &  19.9 & 1.4 &  19.9  \\
     &     &    	     & [S~{\sc iii}]  &  9069  &   4.8 & 0.9 &   4.8  \\
\hline 
\end{tabular}
}
\end{minipage}
\label{tabPN_flux}
\end{table}

\label{lastpage}


\begin{thebibliography}{}

\bibitem[\protect\citeauthoryear{Abazajian et al.}{2009}]{}
Abazajian K. N., Adelman-McCarthy J. K., Ag\"ueros M. A., Allam S. S., \& Allende-Prieto C., et al., 
 2009, ApJS, 182, 543

\bibitem[\protect\citeauthoryear{Andrews \& Martini}{2013}]{}
Andrews B. H., \& Martini P.,  2013, ApJ 765, 140

\bibitem[\protect\citeauthoryear{Baade}{1944}]{}
Baade W., 1944, ApJ, 100, 137 

\bibitem[\protect\citeauthoryear{Balick et al.}{2013}]{}
Balick B., Kwitter K. B., Corradi R. L. M., Henry R. B. C., 2013, ApJ, 774, 3

\bibitem[\protect\citeauthoryear{Belczy\'nski et al.}{2000}]{}
Belczy\'nski K., Miko\'lajewska J., Munari U., Ivison R. J., \& FriedjungX M., 
2000, A\&AS, 146, 407

\bibitem[\protect\citeauthoryear{Benjamin et al.}{1999}]{} 
Benjamin R. A., Skillman E. D., Smits D. P., 1999, ApJ, 514, 307

\bibitem[\protect\citeauthoryear{Bertola et al.}{1995}]{}
Bertola F., Bressan A., Burstein D., Buson 
L.~M., Chiosi C., di Serego Alighieri S., 1995, ApJ, 438, 680 

\bibitem[\protect\citeauthoryear{Bica, Alloin, \& Schmidt}{1990}]{} 
Bica E., Alloin D., Schmidt A.~A., 1990, A\&A, 228, 23 

\bibitem[\protect\citeauthoryear{Cappellari et al.}{1999}]{} 
Cappellari M., Bertola F., Burstein D., 
Buson L.~M., Greggio L., Renzini A., 1999, ApJ, 515, L17 

\bibitem[\protect\citeauthoryear{Cepa \& Beckman}{1988}]{} 
Cepa J., Beckman J.~E., 1988, A\&A, 200, 21 

\bibitem[\protect\citeauthoryear{Ciardullo et al.}{1989}]{} 
Ciardullo R., Jacoby G. H., Ford
H., \& Neill J.D., 1989, ApJ, 339, 53

\bibitem[\protect\citeauthoryear{Clegg}{1987}]{}
Clegg R. E. S., 1987, MNRAS, 229, 31

\bibitem[\protect\citeauthoryear{Corradi et al.}{2005}]{} 
Corradi R.~L.~M., et al., 2005, A\&A, 431, 555 

\bibitem[\protect\citeauthoryear{Davidge}{2003}]{} 
Davidge T.~J., 2003, ApJ, 597, 289 

\bibitem[\protect\citeauthoryear{De Looze et al.}{2012}]{} 
De Looze I., et al., 2012, MNRAS, 423, 
2359 

\bibitem[\protect\citeauthoryear{Ford, Jacoby \& Jenner}{1977}]{}
Ford H. C., Jacoby G., \& Jenner D. C., 1977, ApJ, 213, 18

\bibitem[\protect\citeauthoryear{Gathier \& Pottasch}{1989}]{}
Gathier R., Pottasch S. R., 1989, A\&A, 209, 369

\bibitem[\protect\citeauthoryear{Gilmore \& Wyse}{1991}]{}
Gilmore G., Wyse R. F., 1991, ApJ, 367, L55

\bibitem[\protect\citeauthoryear{Gon{\c c}alves et al.}{2007}]{} 
Gon{\c c}alves D.~R., Magrini L., Leisy 
P., Corradi R.~L.~M., 2007, MNRAS, 375, 715 

\bibitem[\protect\citeauthoryear{Gon{\c c}alves et al.}{2012}]{} 
Gon{\c c}alves D.~R., Magrini L., Martins 
L.~P., Teodorescu A.~M., Quireza C., 2012, MNRAS, 419, 854 

\bibitem[\protect\citeauthoryear{Grebel et al.}{2003}]{}
Grebel E. K., Gallagher J. S., III, \& Harbeck D., 2003, AJ, 125, 1926

\bibitem[\protect\citeauthoryear{Grebel}{2005}]{} 
Grebel E. K., 2005, in Mikolajewska J., Olech A., eds, AIP Conf. Proc.
752, Stellar Astrophysics with the World’s Largest Telescopes: First
International Workshop on Stellar Astrophysics with the World’s Largest
Telescopes. American Institute of Physics, New York, p. 161

\bibitem[\protect\citeauthoryear{Hern\'andez-Mart\'inez et al.}{2009}]{}
Hern\'andez-Mart\'inez L., Pe\~na M., Carigi L., \& Garc\'ia-Rojas J., 2009, A\&A, 505, 1027

\bibitem[\protect\citeauthoryear{Henry}{1989}]{}
Henry R. B. C.,  1989, MNRAS, 241, 453

\bibitem[\protect\citeauthoryear{Hodge}{1973}]{} 
Hodge P.~W., 1973, ApJ, 182, 671 

\bibitem[\protect\citeauthoryear{Howley et al.}{2008}]{}
Howley K. M., Geha M., Guhathakurta P., Montgomery R. M., Laughlin G., \& Johnston K. V.,
2008, ApJ, 683, 722

\bibitem[\protect\citeauthoryear{Izotov et al.}{2006}]{}
Izotov Y. I., Stasi\'nska G., Meynet G., Guseva N. G., \& Thuan T. X., 
2006, A\&A, 448, 955

\bibitem[\protect\citeauthoryear{Jacobi \& Ciardullo}{1999}]{}
Jacoby G. H., \& Ciardullo R., 1999, ApJ, 515, 169

\bibitem[\protect\citeauthoryear{Kaler}{1986}]{}
Kaler J. B., 1986, ApJ, 308, 322

\bibitem[\protect\citeauthoryear{Kaler \& Jacoby}{1989}]{}
Kaler J. B., Jacoby G. H., 1989, ApJ, 345, 871

\bibitem[\protect\citeauthoryear{Kingsburgh \& Barlow}{1994}]{}
Kingsburgh R. L. \& Barlow M. J.,  1994, MNRAS, 271, 257

\bibitem[\protect\citeauthoryear{Kirby et al.}{2013}]{} 
Kirby E. N., Cohen  J. G., Guhathakurta  P., Cheng  L., Bullock  J. S., Gallazzi  A., 2013, ApJ, 779, 102

\bibitem[\protect\citeauthoryear{Kniazev et al.}{2005}]{}
Kniazev A. Y., Grebel E. K., Pustilnik S. A., Pramskij A. G., \& Zucker D. B., 
2005, AJ, 130, 1558

\bibitem[\protect\citeauthoryear{Kniazev et al.}{2007}]{}
Kniazev A. Y., Grebel E. K., Pustilnik S. A., \& Pramskij A. G., 2007, A\&A, 468, 121

\bibitem[\protect\citeauthoryear{Koleva et al.}{2013}]{}
Koleva M., Bouchard A., Prugniel P., De Rijcke S., \& Vauglin I., 2013, MNRAS, 428, 2949

\bibitem[\protect\citeauthoryear{Kormendy \& Djorgovski}{1989)}]{}
Kormendy J, Djorgovski S., 1989, ARA\&A, 27, 235

\bibitem[\protect\citeauthoryear{Krabbe \& Copetti}{2005}]{}
Krabbe A. C. \& Copetti M. V. F., 2005, A\&A, 443, 981

\bibitem[\protect\citeauthoryear{Kwitter et al.}{2012}]{}
Kwitter K. B., Lehman E. M. M., Balick B., \& Henry R. B. C., 2012, ApJ, 753, 12

\bibitem[\protect\citeauthoryear{Lee et al.}{2003}]{}
Lee H., McCall M. L., Kingsburgh R. L., Ross R., Stevenson C. C., 
2003, AJ, 125, 146

\bibitem[\protect\citeauthoryear{Lee et al.}{2008}]{}
Lee H., Bell E. F., \& Somerville R. S., 2008, in IAU Symp. 255, Low-
Metallicity Star Formation: From the First Stars to Dwarf Galaxies, ed. L.
K. Hunt, S. Madden, \& R. Schneider (Cambridge: Cambridge Univ. Press),
100

\bibitem[\protect\citeauthoryear{Leisy \& Dennefeld}{2006}]{} 
Leisy P., Dennefeld M., 2006, A\&A, 456, 451

\bibitem[\protect\citeauthoryear{Leisy et al.}{2005}]{} 
Leisy P., Magrini L., Corradi R., Mampaso A., 2005, AIPC, 804, 265 

\bibitem[\protect\citeauthoryear{Leisy et al.}{2006}]{} 
Leisy P., Magrini L., Corradi R.~L.~M., Mampaso A., 2006, pnbm.conf, 252 

\bibitem[\protect\citeauthoryear{Lianou et al.}{2011}]{}
Lianou S., Grebel E. K., \& Koch A., 2011, A\&A, 531, 152

\bibitem[\protect\citeauthoryear{Magrini et al.}{2005}]{} 
Magrini L., Leisy P., Corradi R.~L.~M., Perinotto M., Mampaso A., Vilchez J.~M., 2005, A\&A, 443, 115 

\bibitem[\protect\citeauthoryear{Magrini \& Gon{\c c}alves}{2009}]{}
Magrini L., Gon{\c c}alves D.~R., 2009, MNRAS, 398, 280 

\bibitem[\protect\citeauthoryear{Magrini et al.}{2009}]{}
Magrini L., Stanghellini L., \&  Villaver E., 2009, ApJ, 696, 729	

\bibitem[\protect\citeauthoryear{Marigo}{2001}]{}
Marigo P., 2001, A\&A, 370, 194

\bibitem[\protect\citeauthoryear{Massey et al.}{1988}]{}
Massey P., Strobel K., Barnes J. V., Anderson E., 1988, ApJ, 328, 315

\bibitem[\protect\citeauthoryear{Massey \& Gronwal}{1990}]{}
Massey P., Gronwall C., 1990, ApJ, 358, 344

\bibitem[\protect\citeauthoryear{Mateo}{1998}]{}
Mateo M. L., 1998, ARA\&A, 36, 435

\bibitem[\protect\citeauthoryear{Mathis}{1990}]{}
Mathis J. S., 1990, ARA\&A, 28, 37

\bibitem[\protect\citeauthoryear{McConnachie}{2012}]{}
McConnachie A. W., 2012, AJ, 144, 4

\bibitem[\protect\citeauthoryear{M\'endez, Kudritzki \& Herrero}{1992}]{}
M\'endez R. H., Kudritzki R. P., \& Herrero A., 1992, A\&A, 260, 329

\bibitem[\protect\citeauthoryear{Monaco et al.}{2009}]{} 
Monaco L., Saviane I., Perina S., Bellazzini M., Buzzoni A., Federici L., Fusi Pecci F., Galleti S., 
2009, A\&A, 502, L9 

\bibitem[\protect\citeauthoryear{Osterbrock \& Ferland}{2006}]{}
Osterbrock D. E. \& Ferland G. J., in Astrophysics of gaseous nebulae and active galactic nuclei, 2nd. ed.
 Sausalito, CA: University Science Books, 2006

\bibitem[\protect\citeauthoryear{Otsuka et al.}{2011}]{}
Otsuka M., Meixner M., Riebel D., Hyung S., Tajitsu A., Izumiura H.,  2011, ApJ, 729, 39

\bibitem[\protect\citeauthoryear{Peimbert \& Torres-Peimbert}{1983}]{}
Peimbert M., Torres-Peimbert S., 1983, in Flower D. R., Proc. IAU Vol. 103,
Planetary Nebulae. Reidel, Dordrecht, p. 233

\bibitem[\protect\citeauthoryear{Pe\~na et al.}{2007}]{}
Pe\~na M., Stasi\'nska G., \& Richer M. G., 2007, A\&A, 476, 745

\bibitem[\protect\citeauthoryear{Perinotto et al.}{2004}]{}
Perinotto M., Morbidelli L., Scatarzi A., 2004, MNRAS, 349, 793

\bibitem[\protect\citeauthoryear{P\'equignot et al.}{2000}]{}
P\'equignot D., Walsh J. R., Zijlstra A. A., \& Dudziak G., 2000, A\&A, 361, L1

\bibitem[\protect\citeauthoryear{Richer \& McCall}{1995}]{}
Richer M. G., McCall M. L., 1995, ApJ, 445, 642

\bibitem[\protect\citeauthoryear{Richer \& McCall}{2007}]{}
Richer M. G., McCall M. L.,  2007, ApJ, 658,328

\bibitem[\protect\citeauthoryear{Richer \& McCall}{2008}]{} 
Richer M. G., McCall M. L., 2008, ApJ, 684, 119

\bibitem[\protect\citeauthoryear{Salaris \& Girardi}{2005}]{}
Salaris M., \& Girardi L., 2005, MNRAS, 357, 669

\bibitem[\protect\citeauthoryear{Saviane et al.}{2000}]{}
Saviane I., Held E. V., \& Bertelli G.,  2000, A\&A, 355, 56

\bibitem[\protect\citeauthoryear{Shaw et al.}{2010}]{}
Shaw R. A., Lee T.-H., Stanghellini L., Davies J. E., Garc\'ia-Hern\'andez D. A., et al., 
2010, ApJ, 717, 562

\bibitem[\protect\citeauthoryear{Stasi\'nska}{1990}]{}
Stasi\'nska G., 1990, A\&AS, 83, 501

\bibitem[\protect\citeauthoryear{Skillman et al.}{1989}]{}
Skillman E. D., Kennicutt R. C., Hodge P. W., 1989, ApJ, 347, 875

\bibitem[\protect\citeauthoryear{Tremonti et al.}{2004}]{}
Tremonti C. A., Heckman T. M., Kauffmann G., Brinchmann J., Charlot S., et al., 
 2004, ApJ, 613, 898

\bibitem[\protect\citeauthoryear{van den Bergh}{2007}]{}
van den Bergh S., 2007, in The Galaxies of the Local Group, Cambridge: Cambridge University Press, 2007

\bibitem[\protect\citeauthoryear{van Zee et al.}{2006}]{}
van Zee L., Skillman E D., \& Haynes M. P., 2006, ApJ, 637, 269

\bibitem[\protect\citeauthoryear{Vassiliadis \& Wood}{1993}]{}
Vassiliadis E., Wood P. R., 1993, ApJ, 413, 641

\bibitem[\protect\citeauthoryear{Vassiliadis \& Wood}{1994}]{}
Vassiliadis E., Wood P. R., 1994, ApJS, 92, 125

\bibitem[\protect\citeauthoryear{Vigroux et al.}{1987}]{}
Vigroux L., Stasi\'nska G., \& Comte G., 1987, A\&A, 172, 15

\bibitem[\protect\citeauthoryear{Young \& Lo}{1997}]{} 
Young L.~M., Lo K.~Y., 1997, ApJ, 476, 127 

\bibitem[\protect\citeauthoryear{Zijlstra \& Pottasch}{1989}]{}
Zijlstra A. A., Pottasch S. R., 1989, A\&A, 216, 245

\bibitem[\protect\citeauthoryear{Zijlstra et al.}{2006}]{}
Zijlstra A. A., Gesicki K., Walsh J. R., P\'equignot D., van Hoof P. A. M., \& Minniti D., 
2006, MNRAS, 369, 875

\end{thebibliography}
\end{document}